\documentclass{aa}  
\usepackage{graphicx}
\usepackage{txfonts}
\begin{document}
   \title{The XMM-{\it Newton} Wide Angle Survey (XWAS): the X-ray spectrum of type-1 AGN}
   \subtitle{}

   \author{S. Mateos \inst{1}
     \and
     F.J. Carrera \inst{2}
     \and
     M.J. Page \inst{3}
     \and
     M.G. Watson  \inst{1}
     \and
     A. Corral \inst{4}
     \and
     J.A. Tedds \inst{1}
     \and 
     J. Ebrero \inst{5}
     \and     
     M. Krumpe \inst{6}
     \and     
     A. Schwope \inst{7}
     \and     
     M.T. Ceballos \inst{2}
          }

   \offprints{S. Mateos, \email{sm279@star.le.ac.uk}}

   \institute{
	  Department of Physics and Astronomy, University of Leicester. University Road, Leicester, UK
	  \and Instituto de F\'\i sica de Cantabria (CSIC-UC), 39005 Santander, Spain
	  \and AA (Mullard Space Science Laboratory, University College London, Holmbury St Mary, Dorking, Surrey RH5 6NT)
	  \and INAF-Osservatorio Astronomico di Brera, via Brera 28, I-20121 Milan, Italy
	  \and SRON - Netherlands Institute for Space Research, Sorbonnelaan 2, 3584 CA, Utrecht, The Netherlands
	  \and University of California, San Diego, Center for Astrophysics \& Space Sciences, 9500 Gilman Drive, La Jolla, CA 92093-0424, USA 
	  \and Astrophysikalisches Institut Potsdam, An der Sternwarte 16, 14482 Potsdam, Germany
             }

   \date{11 December 2009}

 
  \abstract
   {}
   {We discuss the broad band X-ray properties of one of the largest samples of X-ray selected type-1 AGN to date (487 objects in total), drawn
     from the XMM-{\it Newton} Wide Angle Survey (XWAS). The objects presented in this work cover 2-10 keV (rest-frame) luminosities 
     from ${\rm \sim10^{42}-10^{45}\,erg\,s^{-1}}$ and are detected up to 
     redshift $\sim$4. We constrain the overall properties of the broad band continuum, soft excess and X-ray absorption, along with their 
     dependence on the X-ray luminosity and redshift. We discuss the implications for models of AGN emission.
     }
   {We fitted the observed 0.2-12 keV broad band spectra with various models to search for X-ray absorption and soft excess. The F-test was used with a significance 
     threshold of 99\% to statistically accept the detection of additional spectral components. 
     }
   {We constrained the mean spectral index of the broad band X-ray continuum to $\langle\Gamma\rangle=1.96\pm0.02$ with intrinsic dispersion 
     ${\rm \sigma_{\langle \Gamma \rangle}=0.27_{-0.02}^{+0.01}}$. The continuum becomes harder at faint fluxes and at higher redshifts and hard (2-10 keV) luminosities.
     The dependence of $\Gamma$ with flux is likely due to undetected absorption rather than to spectral variation. 
     We found a strong dependence of the detection efficiency of objects on the spectral shape.
     We expect this effect to have an impact on the measured mean continuum shapes of sources at different redshifts
     and luminosities. We detected excess absorption in $\gtrsim$3\% of our objects, with rest-frame column densities $\sim$a few ${\rm \times10^{22}\,cm^{-2}}$. The apparent mismatch between the optical classification 
     and X-ray properties of these objects is a challenge for the standard orientation-based AGN unification model.
     We found that the fraction of objects with detected soft excess is $\sim$36\%. Using a thermal model, we constrained the soft excess 
     mean rest-frame temperature and intrinsic dispersion to {\it kT}$\sim$100 eV and $\sigma_{\it kT}$$\sim$34 eV.
     The origin of the soft excess as thermal emission from the accretion disk or Compton scattered 
     disk emission is ruled out on the basis of the temperatures detected and the lack of correlation of the soft excess 
     temperature with the hard X-ray luminosity over more than 2 orders of magnitude in luminosity. Furthermore, the high luminosities of the soft 
     excess rule out an origin in the host galaxy. }
   {}


   \maketitle
%

\section{Introduction}
It is now widely accepted that the emission of Active Galactic Nuclei (AGN) in the X-ray band originates in the innermost 
regions, close to the energy source of the AGN, most probably an accreting 
supermassive black hole (SMBH). The study of the X-ray emission of AGN 
provides strong constraints on their powering mechanism and gives insight  
to the geometry and physical conditions of the matter surrounding the SMBH. 

The typical spectrum of both low luminosity Seyfert galaxies 
and bright QSOs in the hard X-ray band ($\gtrsim$2 keV) is dominated by a power-law emission with a 
photon index $\Gamma$$\sim$2 with intrinsic dispersion 
$\sigma_{\Gamma}$$\sim$0.2-0.3 (e.g. Mainieri et al.~\cite{Mainieri07}; Tozzi et al.~\cite{Tozzi06}; Mateos et al.~\cite{Mateos05b}). 
This hard X-ray spectral component is believed to originate in hot plasma surrounding 
the accretion disk which Compton up-scatters UV-soft X-ray thermal emission from the disk into the 
hard X-ray band (Haardt \& Maraschi~\cite{Haardt91}). 

Part of the primary power-law continuum emission is then reprocessed in the disk producing 
a reflection spectrum characterised by a hardening of the spectrum above $\sim$10 keV and a fluorescent 
Fe K$\alpha$ emission line around $\sim$6.4 keV mostly detected in Seyfert-like luminosity objects (Pounds et al.~\cite{Pounds90}). 
Signatures of ionised and/or cold gas have also been detected in high signal-to-noise X-ray spectra of both 
Seyfert galaxies and QSOs (e.g. Piconcelli et al.~\cite{Piconcelli05}). Moreover, an important fraction of AGN shows an excess of emission with respect to 
the power-law emerging below $\sim$2 keV, the so-called {\it soft excess}, whose nature is still debated. 
The soft excess has been interpreted as the tail of the thermal emission from the accretion 
disk (Turner \& Pounds~\cite{Turner89}) or Comptonization of EUV accretion disk 
photons (Kawaguchi et al.~\cite{Kawaguchi01}). However, when the soft excess is fitted with a thermal model, the measured temperature is roughly constant over several orders 
of magnitude in luminosity and BH mass ($\langle\,{\it kT}\,\rangle$$\sim$0.1 keV, Crummy et al.~\cite{Crummy06}; 
Gierli\'nski \& Done~\cite{Gierlinski04}). This is a major problem for models based on accretion disk continuum emission
because the disk temperature is expected to vary with both the mass of the black hole and 
the accretion rate. Furthermore, the measured soft excess temperatures are too high to be the high energy tail of the 
accretion disk emission. In the past few years two alternative models with very different geometries have been 
proposed to explain the origin of the soft excess, both relating this spectral component to atomic 
rather than continuum processes: the soft excess could be produced by absorption from a relativistically outflowing 
warm gas which is optically thin and in the line of sight, i.e. a wind above the 
disk (Gierli\'nski \& Done~\cite{Gierlinski04}), or by enhancement of reprocessing of disk emission in optically thick 
material out of the line of sight (Crummy et al.~\cite{Crummy06}). These two models are indistinguishable in the 0.2-10 keV 
band (see Ponti et al.~\cite{Ponti08} and references therein).

   \begin{figure*}[!t]
   \centering
   \includegraphics[angle=90,width=0.49\textwidth]{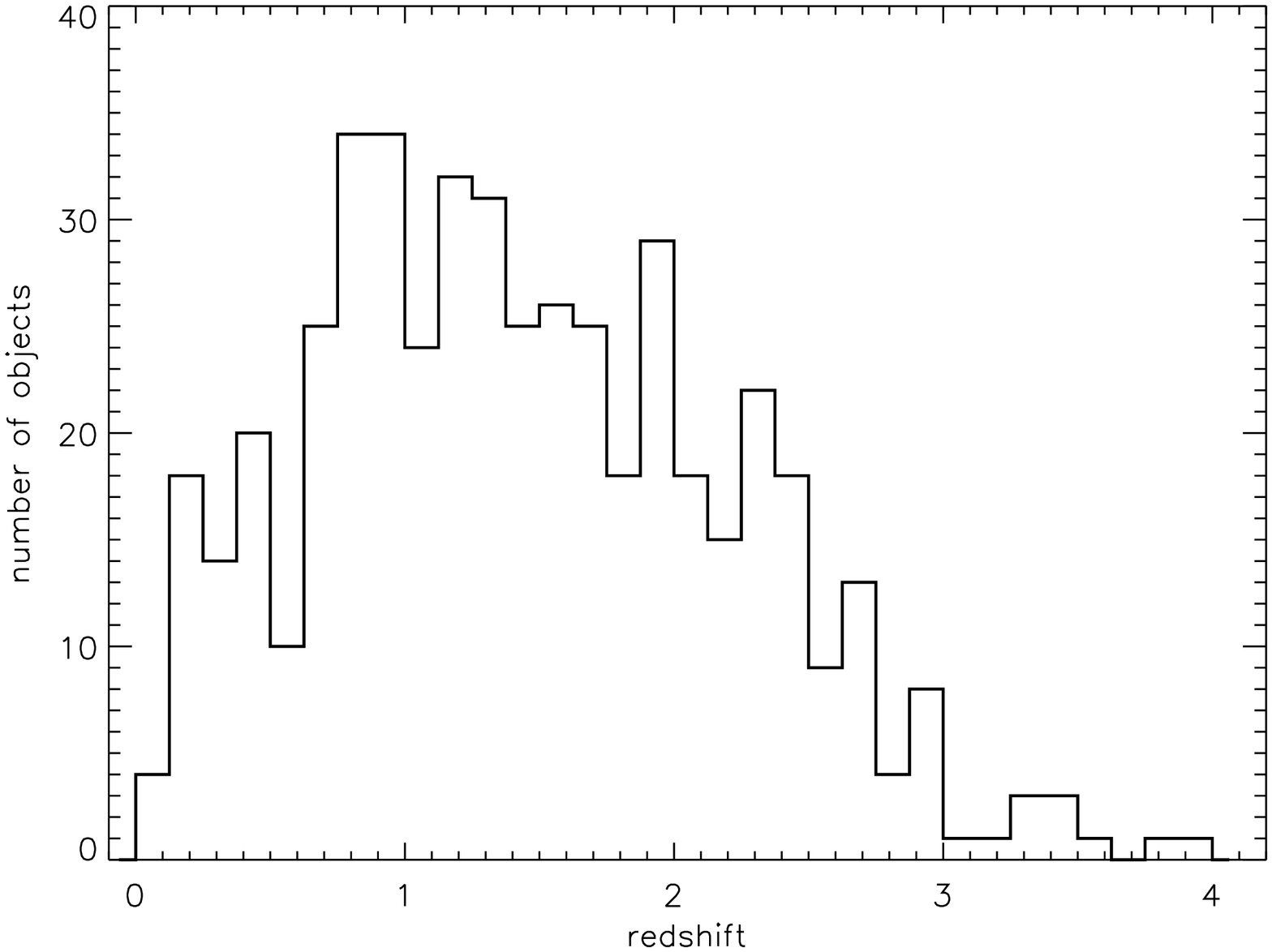}
   \includegraphics[angle=90,width=0.49\textwidth]{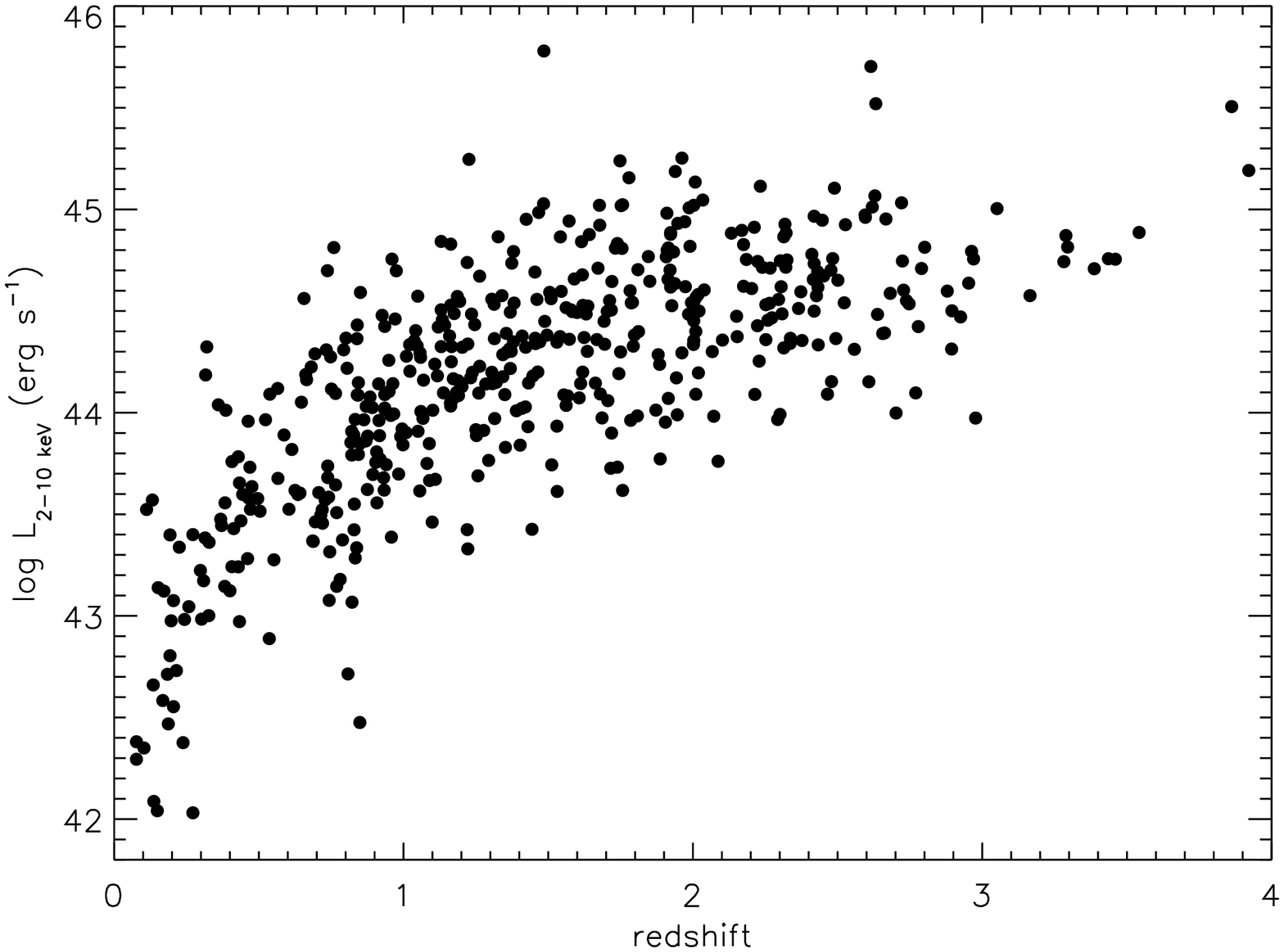}
   \hbox{
   \includegraphics[angle=90,width=0.49\textwidth]{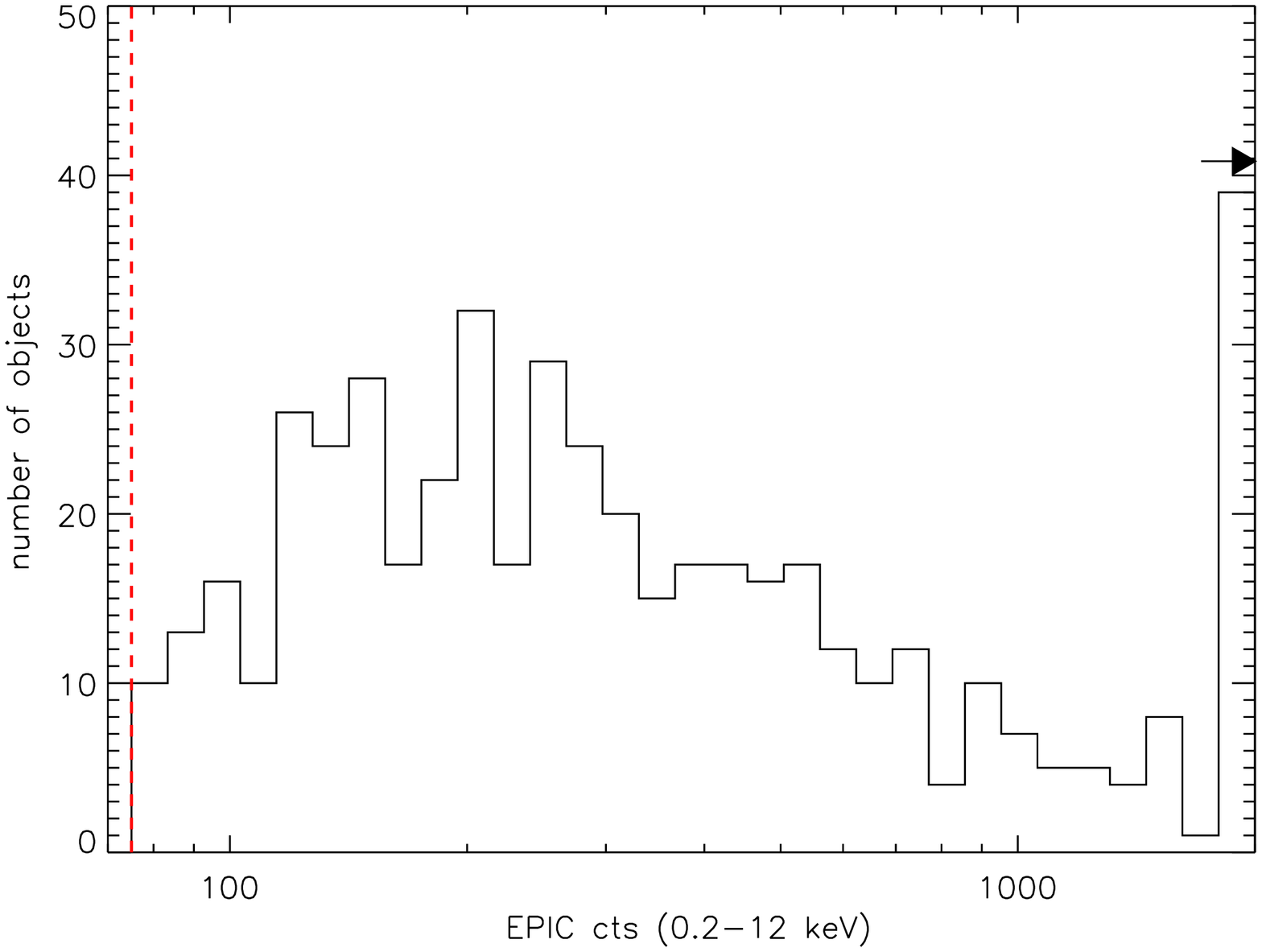}
   \includegraphics[angle=90,width=0.49\textwidth]{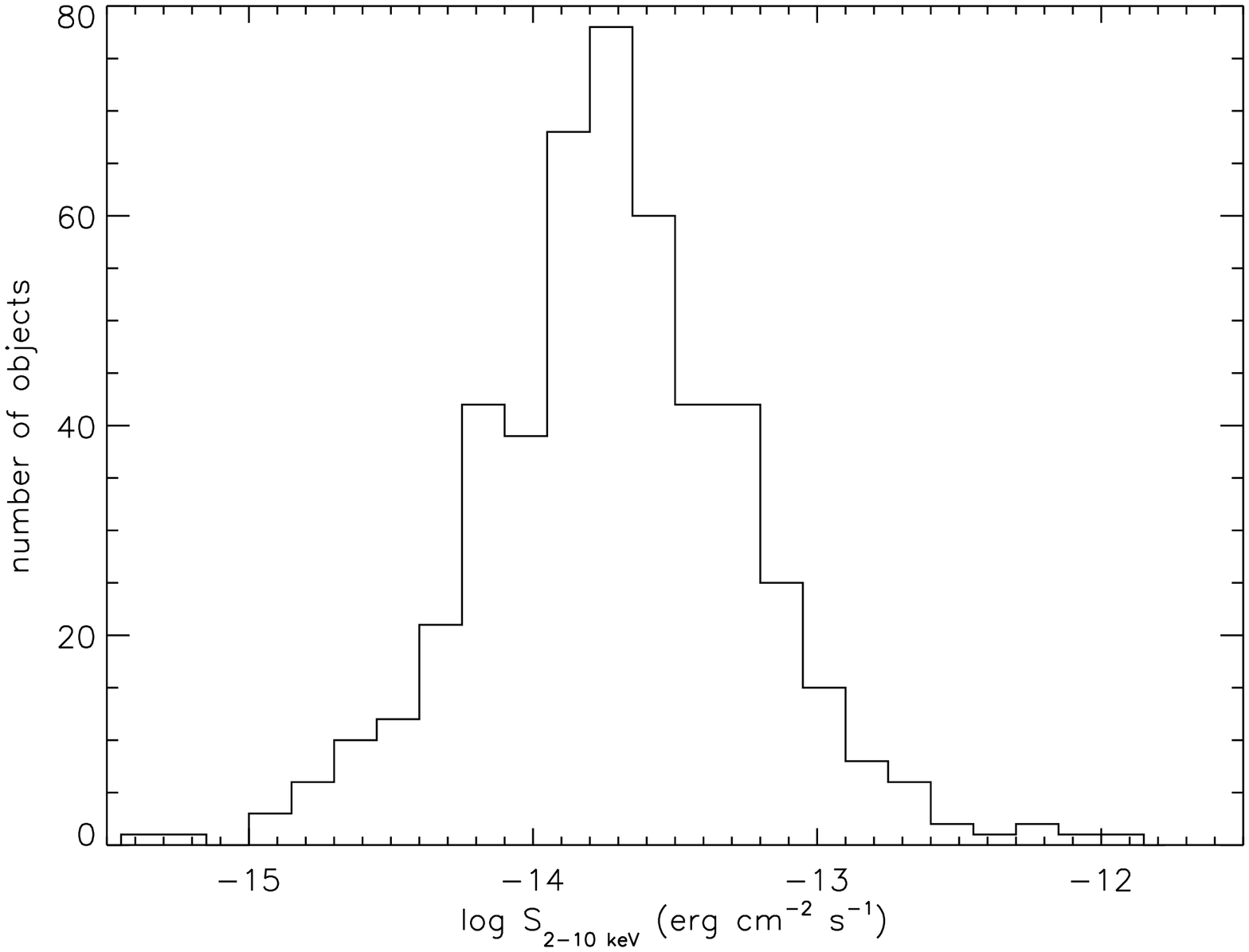}}
   \caption{Top: Redshift distribution (left) and 2-10 keV X-ray 
     luminosity (corrected for absorption) vs. redshift (right) of the XWAS type-1 AGN selected for our study.
     Bottom: Distribution of EPIC 0.2-12 keV net counts in the X-ray spectra of the sources (left)
     and 2-10 keV band flux distribution (right). For clarity the last bin in the distribution of EPIC counts includes all
     objects with spectra with more than 2000 counts. The dashed vertical line indicates the limit in counts we used for this study.
     }
              \label{z_dist}%
    \end{figure*}

In most surveys AGN are optically classified as type-1 AGN if broad lines with velocity widths ${\rm \gtrsim 1500\,km\,s^{-1}}$ 
are present in their optical spectra. Objects where only narrow lines (velocity widths ${\rm \lesssim 1500\,km\,s^{-1}}$) 
are seen are classified as type-2 AGN. According to AGN unification models the different properties observed for 
type-1 and type-2 AGN can be explained solely as an orientation effect (e.g. Antonucci~\cite{Antonucci93}; Urry \& Padovani~\cite{Urry95}). 
In type-1 AGN our line of sight to the nucleus is not intercepted by optically 
obscuring material, and therefore we see the central engine and the broad line region. The line of sight in type-2 AGN intercepts the obscuring 
material hiding both the central engine and the broad line region. 
Therefore type-1 AGN are expected to show little or no obscuration in X-rays while type-2 AGN should be more heavily absorbed in X-rays. Spectral 
analyses have shown that the optical type vs. X-ray properties for most AGN generally agree with the predictions from AGN unification models.
However, in recent years, the number of cases where the predictions from AGN unification models do not match the observations has increased 
substantially, as significant absorption has been detected in $\sim$10\% type-1 AGN 
(Cappi et al.~\cite{Cappi06}; Mateos et al.~\cite{Mateos05b}; Akylas et al.~\cite{Akylas04}) 
while type-2 AGN have been detected without any signatures of X-ray absorption (Panessa et al.~\cite{Panessa09}; Bianchi et al.~\cite{Bianchi08}; 
Corral et al.~\cite{Corral05}; Panessa \& Bassani~\cite{Panessa02}).

It is still not clear whether the X-ray properties of AGN depend on the luminosity or have any spectral variation with redshift. Various studies of the 
X-ray photon index have produced different results. Some authors claim an anti-correlation between the spectral photon index $\Gamma$ and 
the X-ray luminosity (Young et al.~\cite{Young09}; Saez et al.~\cite{Saez08}; Page et al.~\cite{Page05}) while other studies suggest the opposite 
correlation (Dai et al.~\cite{Dai04}) or no correlation at 
all (Winter et al.~\cite{Winter09}; Mateos et al.~\cite{Mateos05b}; George et al.~\cite{George00}; Reeves \& Turner~\cite{Reeves00}). Studies of 
the X-ray spectra of AGN have generally shown no clear dependence of the AGN X-ray properties on the 
redshift (Mateos et al.~\cite{Mateos05a}), although some exceptions have been found (Kelly et al.~\cite{Kelly07}).

Here we investigate the broad band X-ray properties of 487 type-1 AGN from the XMM-{\it Newton} Wide Angle 
Survey (XWAS). We selected only those objects having X-ray spectra with sufficient signal-to-noise ratio (hereafter S/N) to perform a basic spectral analysis. 
This is one of the largest X-ray selected samples of type-1 AGN assembled to date for which constraints on the X-ray properties of the sources 
can be directly derived from a spectral analysis. A study of the X-ray properties of 156 type-1 AGN observed by XMM-{\it Newton} was recently presented 
by Bianchi et al.~(\cite{Bianchi08}). As this study is based on type-1 AGN in targeted observations, it has the advantage that most of the spectra 
are characterised by high S/N, but it has the disadvantage of complex selection function, biased towards 'interesting' objects, and it mainly 
samples AGN in the local Universe. The type-1 AGN from XWAS span more than 3 orders of magnitude in X-ray luminosity and 
are detected at redshifts from 0.08 to 3.9. Hence, our type-1 AGN sample is ideal to investigate whether the properties of the broad band 
continuum shape, X-ray absorption and soft excess in type-1 AGN depend on the X-ray luminosity of the objects and whether they have evolved 
with cosmic time.

This paper is organised as follows: In \textsection2 we describe the XMM-{\it Newton} Wide Angle Survey and the compilation of the sample of type-1 AGN.
In \textsection3 we discuss the procedure for extraction of the X-ray spectra of the objects and the spectral fitting. 
In \textsection4  we present and discuss the results of the spectral analysis. The summary and conclusions of our analysis are reported in \textsection5. 
Throughout this paper we have adopted the WMAP derived cosmology with $H_0=70\,{\rm km\,s^{-1}\,Mpc^{-1}}$, $\Omega_{\rm M}=0.3$ and 
$\Omega_{\Lambda}=0.7$ (Spergel et al.~\cite{Spergel03}). 

\section{The XMM-{\it Newton} Wide Angle Survey (XWAS)} 
\label{xwas}
The XMM-Newton Wide Angle Survey (XWAS) identification programme was based on optical multi-fibre spectroscopy obtained at the Anglo Australian Telescope
(AAT) Two Degree Field (2dF; Lewis et al.~\cite{Lewis02}) for the counterparts of serendipitous X-ray sources,
originally selected from 68 spatially distinct pointings made by XMM-{\it Newton} between June 2000 and May 2003. The fields  targeted by XWAS cover a net 
sky area $\Omega$$\sim$11.5 deg$^2$. A full description of this survey will be presented in a forthcoming paper.

The XMM-Newton field selection for XWAS prioritised those observations with adjacent or overlapping coverage, to take optimum advantage of the 2 degree
diameter of the 2dF spectrograph. Counterparts for 2dF spectroscopy were primarily selected from the SuperCOSMOS Sky Survey (Hambly et al.~\cite{Hambly01}),
supplemented by additional counterparts extracted from Isaac Newton Telescope Wide Field Camera (INT WFC) imaging. Only candidates down 
to R$\approx$21 were selected, with priority given to X-ray sources with 0.5-4.5 keV flux $\geq 10^{-14}\rm\ erg\ cm^{-2}\ s^{-1}$. The XWAS programme 
was primarily aimed at accumulating as large a sample of X-ray sources identifications as possible and is not complete to these magnitude and X-ray flux limits.

Optical spectroscopy was obtained for the potential counterparts of a total of $\sim$3000 X-ray sources (although, as the 2dF provides more fibres per field than
required for this programme, a significant fraction of the spectroscopic fibres were placed on lower probability counterparts and X-ray sources 
with lower detection likelihoods). The XWAS 2dF spectroscopy provides an effective resolution $\lambda/\delta\lambda\approx 600$ over a wavelength range 
$\sim$3850-8250 \AA\ and 
reaches a S/N of $\sim$5 at 5500 \AA\ for V=21 mag, sufficient to provide reliable object classification and redshift determination together with a reasonable 
characterisation of the optical continuum shape.

The XWAS survey provides data of sufficient quality for the optical spectroscopic identification of $\approx$980 of the sample X-ray sources. Of these,
the majority, $\approx$65\%, are identified as type-1 AGN or BLAGN (velocity widths $\geq$1500 km s$^{-1}$). The remainder comprise  narrow emission line 
galaxies (NELGs; $\approx$16\%, velocity widths $\leq$1500 km s$^{-1}$) absorption line galaxies ($\approx$6\%) and Galactic stars ($\approx$13\%).

Corral et al.~(\cite{Corral08}) computed the rest-frame X-ray averaged spectrum and constrained the mean Fe K$\alpha$ line emission properties for type-1 AGN up
to redshift $\sim$3.5 while Krumpe et al.~(\cite{Krumpe08}) investigated the X-ray properties of the most luminous type-2 AGNs. In this paper we concentrate on the X-ray 
spectral analysis of the 487 type-1 AGN in the sample with sufficient S/N in their X-ray spectra 
($\ge$75 EPIC counts in the 0.2-12 keV energy band) to provide constraints on their broad band X-ray properties. The top row 
in Fig.~\ref{z_dist} shows the redshift distribution of the sources (left) and the distribution of 2-10 keV luminosity vs. redshift (right). 
The bottom  row shows the distributions of 0.2-12 keV EPIC net 
counts (left) and 2-10 keV fluxes (right). Fluxes and luminosities were computed from the best-fit model and 
corrected for Galactic absorption. Luminosities are also corrected for intrinsic X-ray absorption.
The great majority of AGN, including type-1 AGN, have X-ray-to-optical flux ratios (X/O) in the range 0.1$<$X/O$<$10 
(e.g. Brusa et al.~\cite{Brusa07}). The imposed limit of R$\sim$21 for optical follow-up of the XWAS sources, implies that as we 
approach the X-ray flux limit of the survey we are better sampling type-1 AGN with typically lower X-ray to optical flux ratios than the overall population. 
We note however that X-ray spectral analyses of AGN samples not suffering from this identification bias 
do not find that the X-ray properties of type-1 AGN vary with X/O (see e.g. Tozzi et al.~\cite{Tozzi06}). 
Hence we do not expect the imposed limit for optical follow-up of the XWAS 
sources to have any impact on the results of our analysis.

We matched our type-1 sample with the 1.4-GHz NRAO VLA Sky Survey (NVSS, Condon et al.~\cite{Condon98}). The survey 
covers the entire sky north to -40 deg declination, hence we have radio coverage for 410 out of the 487 objects in our sample. Only 10 objects 
were detected above the NVSS completeness limit of 2.5 mJy. Therefore we expect the contribution to our sample from radio-loud AGN to be less than 
3\%, and hence our sample to be dominated by radio quiet AGN.

\subsection{The X-ray data}
\label{xwas_xdata}
We searched the XMM-{\it Newton} archive for all the available observations of XWAS sources. The number of 
XMM-{\it Newton} fields that overlap with the original XWAS selected pointings at the time of this study was 119 observations in total.

In order to get the X-ray source list for each observation we used the same source detection 
configuration as in the second XMM-{\it Newton} serendipitous source catalogue, {\tt 2XMM} (see Watson et al.~\cite{Watson08} for details). 
Source detection is carried out on data of the three EPIC cameras (pn, MOS1 and MOS2) and on five 
energy bands simultaneously (0.2-0.5 keV, 0.5-1 keV, 1-2 keV, 2-4.5 keV and 4.5-12 keV). 
The final XWAS X-ray source catalogue includes all detections with EPIC 0.2-12 keV detection likelihood $\ge$6 and 
contains $\sim$7000 unique X-ray sources.

   \begin{figure}[!t]
   \centering
   \hbox{
   \includegraphics[angle=90,width=0.5\textwidth]{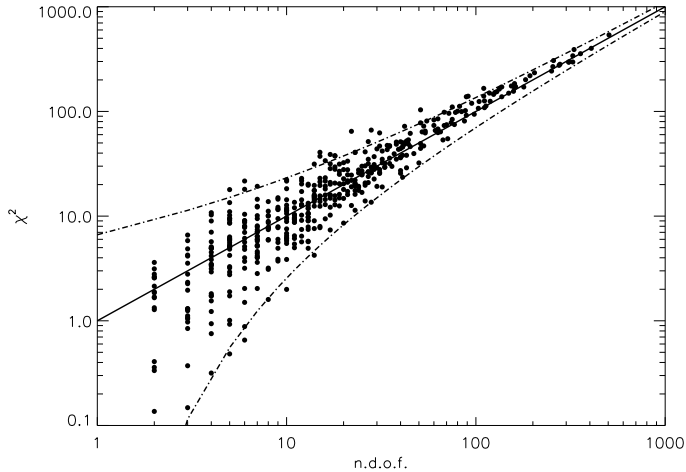}}
   \caption{Best-fit $\chi^2$ vs. ndof of the fits (circles). The solid line indicates the value corresponding 
     to $\chi^2$=ndof. The upper and lower dashed lines indicate for a given number of ndof the $\chi^2$ range above 
     and below which we expect a 1\% probability of finding such high/low values of $\chi^2$ if the model is correct.
     }
              \label{chi2_vs_dof}%
    \end{figure}

\section{Spectral analysis} 

\subsection{X-ray spectral extraction}
Spectra were extracted for each detection and EPIC camera in circular regions centred at the source positions. 
The size of the source extraction regions was selected to optimise the S/N via the {\tt SAS} task {\tt eregionanalyse}. Background 
spectra were obtained in annular regions centred at the source position after masking out nearby sources falling in the extraction 
regions. If the background region did not have sufficient S/N we used source-free circular regions near the sources and in the same CCD. Appropriate response matrices and effective area curves were obtained for each detection 
using the {\tt SAS} tasks {\tt rmfgen} and {\tt arfgen} respectively.

For each source all detections were merged to obtain a pn and a single MOS spectrum over the observed 
energy range 0.2-12 keV. As the differences between the MOS1 and MOS2 response matrices are 
of just a few per cent we created combined MOS source and background spectra and response matrices. 
Merged source and background spectra were obtained by adding the individual spectra. 
Backscale values (size of the regions used to extract the spectra) and calibration matrices for the combined 
spectra were obtained weighting the input data with the exposure times. 
In order to use the $\chi^2$ minimisation during the spectral fitting spectra were grouped 
to a minimum of 15 counts per bin. 

\subsection{Spectral fitting}
\label{sp_fitting}
The spectral fitting of the 487 XWAS type-1 AGN with more than 75 EPIC counts in the 0.2-12 keV band 
was carried out with the XSPEC package (Arnaud~\cite{Arnaud96}). Hereafter 
quoted errors refer to the 90\% confidence level for one interesting parameter (i.e. $\Delta\chi^2$=2.71) unless otherwise stated.
We started with a joint fitting of MOS and pn spectra with a power-law model with fixed Galactic absorption ($phabs \times powerlaw$ in XSPEC). The spectral 
parameters were tied to the same value while the normalisations were left free to vary in order to account for flux cross-calibration discrepancies between the EPIC 
MOS and pn cameras (see Mateos et al.~\cite{Mateos09}). The Galaxy H column density values for each field were obtained using the 
ftool {\it nh}\footnote{http://heasarc.nasa.gov/lheasoft/ftools/}. The values are derived from the HI map of Dickey \& Lockman~(\cite{Dickey90}). 
Two additional model components were added to the simple power-law model to search for 
intrinsic absorption and soft excess emission: a redshifted neutral absorption component ($zphabs$ in XSPEC) and a blackbody 
spectrum ($zbbody$ in XSPEC). We also used a partial covering model ($zpcfabs$ in XSPEC) to fit the spectral complexity observed in 
some of our sources, but we found that the quality of the fits did not improve significantly for any source with respect to the above models.

The significance of the detection of additional components in the X-ray spectra of our 
sources was measured with the F-test. The F-test measures the significance of a change in $\chi^2$ when new components are 
added to the model. We used a significance threshold of 99\% to accept the detection of soft excess and/or 
intrinsic absorption. No more spectral complexity was detected, hence there was no need for additional model components. 
The lack of detection of the Fe K$\alpha$ line emission 
is likely due to the fact that most spectra do not have sufficient S/N to detect this spectral component. However, it has also been shown that 
the most luminous AGN do not tend to reveal iron emission (e.g. Corral et al.~\cite{Corral08}; Krumpe et al.~\cite{Krumpe08}; Iwasawa \& Taniguchi~\cite{Iwasawa93}). 

   \begin{figure}[!t]
   \centering
   \hbox{
   \includegraphics[angle=90,width=0.49\textwidth]{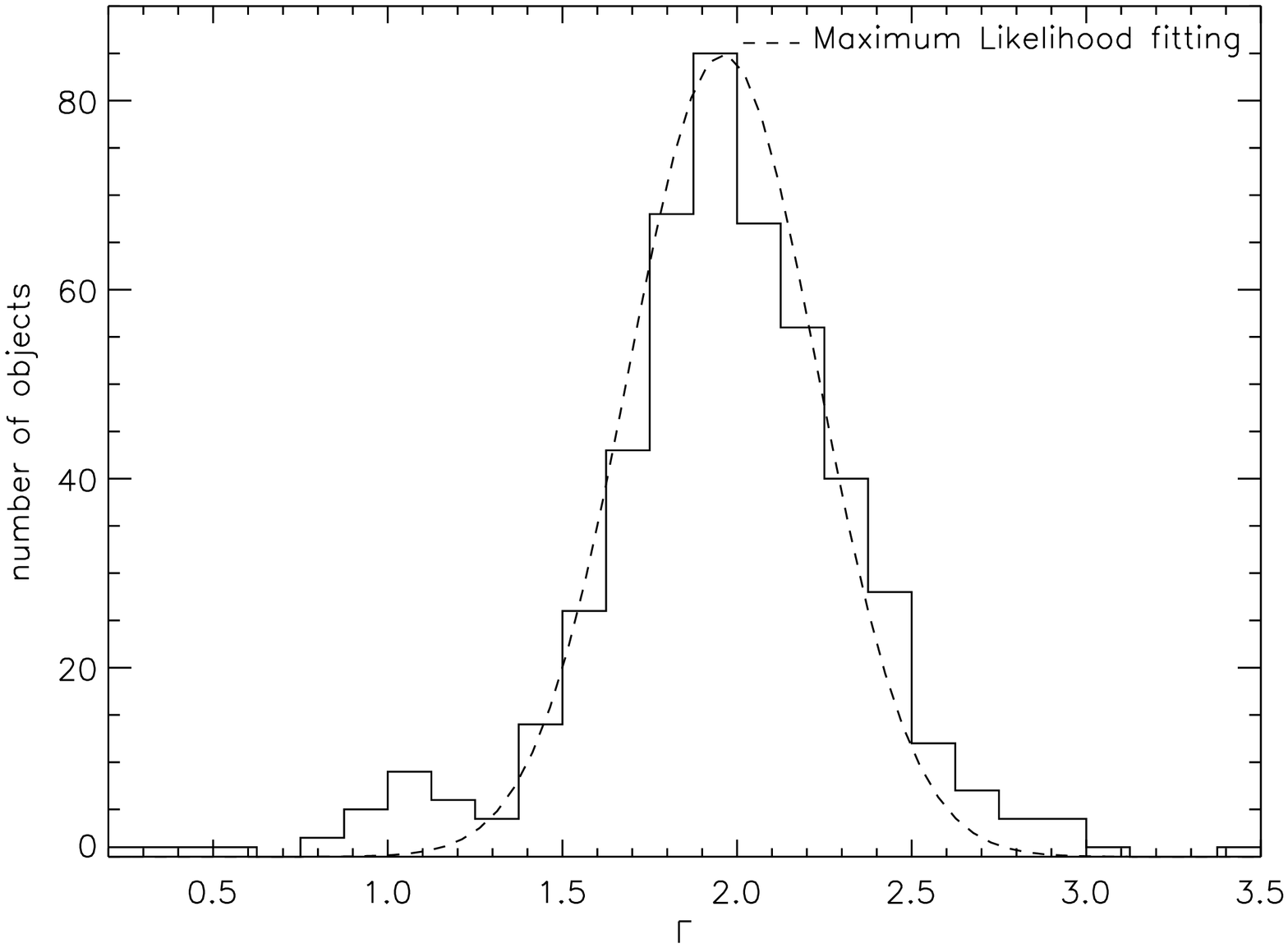}}
   \hbox{
   \includegraphics[angle=90,width=0.49\textwidth]{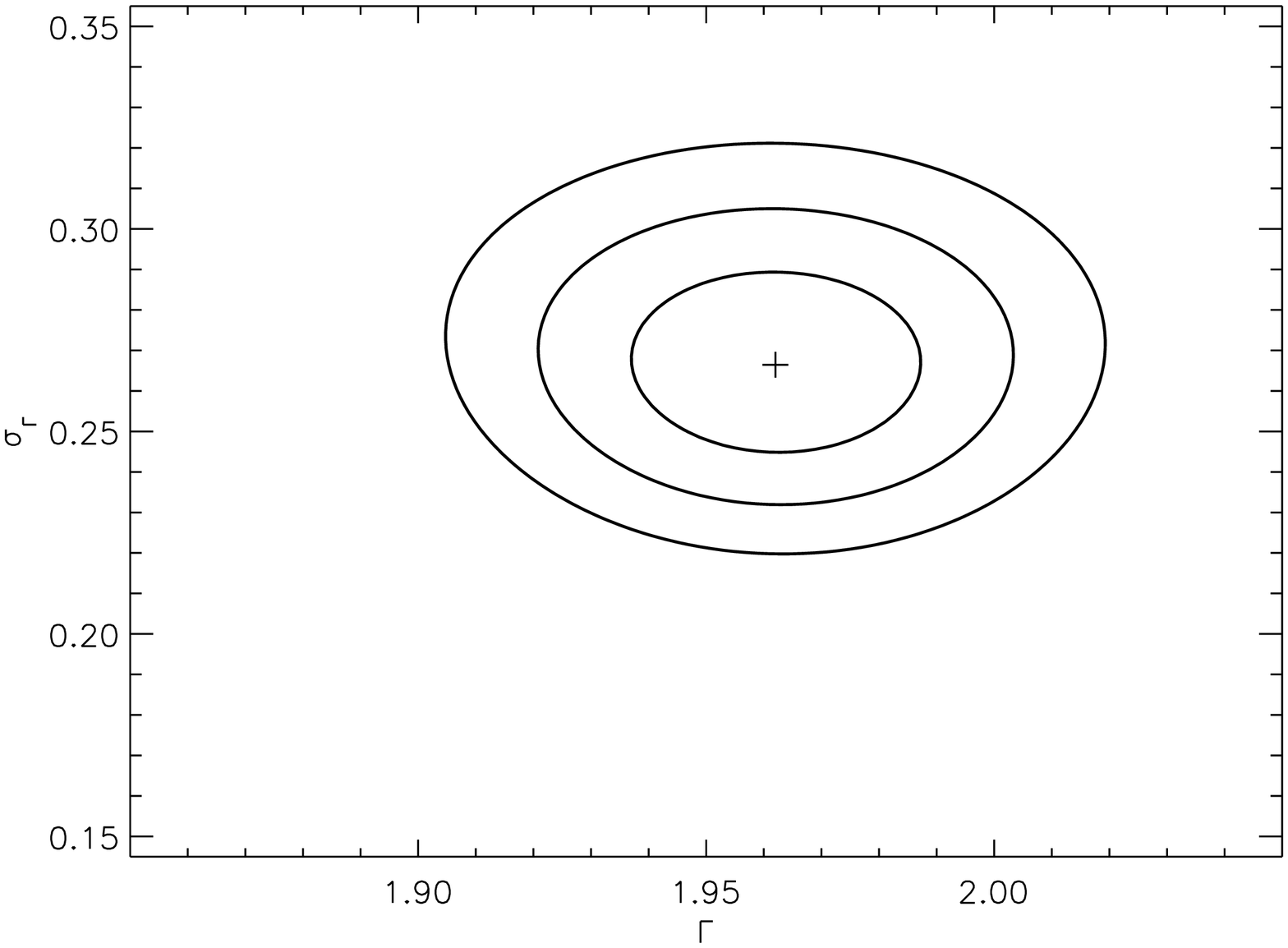}}
   \caption{Top: Distribution of best-fit photon indexes. The dashed line shows the results of the maximum likelihood technique 
     used to constrain the mean spectral index of the broad band X-ray continuum and its intrinsic dispersion (see text for details).
     Bottom: 1$\sigma$, 2$\sigma$ and 3$\sigma$ contours for the photon index $\Gamma$
     and its intrinsic dispersion, $\sigma_{\Gamma}$. The cross indicates 
       the best-fit values. 
   }
     \label{gmm_dist}%
    \end{figure}

Fig.~\ref{chi2_vs_dof} shows the quality of the spectral fitting, where 
the best-fit $\chi^2$ values are shown as a function of the number of degrees of freedom (ndof) of the fits. 
The solid line indicates the value corresponding to $\chi^2$=ndof. The upper and lower dashed lines 
indicate for a given number of ndof the range of $\chi^2$ above and below which we expect a 1\% probability of finding such high/low 
values of $\chi^2$ if the model is correct. We see that the quality of our fits is very good, only 5\% and 1\% of the fits 
have $\chi^2$ values that lie above or below the region delineated by the dashed lines respectively.

\begin{table}[!t]
  \caption{Summary of the results of the spectral fitting.}
\label{table:1}      
\centering                          
\begin{tabular}{c c c c c c c c c c c c c c c c c c }        
\hline\hline                 
Model &  ${\rm N_{total}}$ \\
(1) &  (2) \\
\hline                        
$phabs \times powerlaw$ &  430 (88.3\%) \\
$phabs \times (zphabs \times powerlaw)$ &   17 (3.5\%)\\
$phabs \times (zbbody + powerlaw)$  &   38 (7.8\%)\\
$phabs \times (zbbody + zphabs \times powerlaw)$  &    2 (0.4\%)\\
\hline                                   
\end{tabular}
\\(1) XSPEC model definition: $powerlaw$: Simple photon power-law. $zphabs$: Rest-frame photoelectric absorption. $zbbody$: A redshifted blackbody spectrum. 
All models include Galactic absorption ($phabs$). 
(2) Number and fraction of objects best fitted with the indicated model.
\end{table}

A summary of the results of the spectral fitting is presented in Table~\ref{table:1}. As expected most of our type-1 
AGN are best fitted with a simple power-law model corrected for Galactic absorption. 
Excess X-ray absorption above the Galactic value and a soft excess emission were detected 
in $\sim$4\% and $\sim$9\% of the sources respectivelly. We used the recipe in Mateos et al.~(\cite{Mateos05a}) to obtain the intrinsic 
fraction of objects with detected soft excess and/or intrinsic absorption. The method 
takes into account the expected number of spurious detections (1\% expected for the chosen F-test significance threshold)
using Bayesian statistics. We found that intrinsic absorption and soft excess emission were detected with a significance 
$\geq$99\% in 3\%$\pm$1.5\% and 7.3\%$\pm$2\% of the sources respectively.

\section{Results and discussion} 
\subsection{The broad band X-ray continuum} 
\label{gamma}
Fig.~\ref{gmm_dist} (top) shows the distribution of best-fit $\Gamma$ values (solid line). In order to constrain the mean continuum shape of 
our type-1 AGN, $\langle \Gamma \rangle$, we have 
assumed that the distribution of spectral slopes can be well reproduced with a Gaussian of mean $\langle \Gamma \rangle$ and 
dispersion ${\rm \sigma_{\langle \Gamma \rangle}}$. The best simultaneous estimates of the average photon 
index $\langle \Gamma \rangle$ and the intrinsic spread ${\rm \sigma_{\langle \Gamma \rangle}}$ 
were obtained with a maximum likelihood technique accounting for both the errors in the measurements and the intrinsic dispersion 
of values (Maccacaro et al.~\cite{Maccacaro88}). Fig.~\ref{gmm_dist} shows the best-fit Gaussian distribution (top, dashed line) 
and the 1$\sigma$, 2$\sigma$ and 3$\sigma$ contours for the 
two parameters together with the best-fit values, $\langle\Gamma\rangle=1.96\pm0.02$ 
and ${\rm \sigma_{\langle \Gamma \rangle}=0.27_{-0.02}^{+0.01}}$ (cross). The errors in the 
parameters were obtained from the 1$\sigma$ contour. These values are in excellent 
agreement with recent measurements of the mean photon index and intrinsic dispersion 
for type-1 AGN found in the literature (e.g. Young et al.~\cite{Young09}; Dadina~\cite{Dadina08}; Mainieri et al.~\cite{Mainieri07}, Page et al.~\cite{Page06}; 
Mateos et al.~\cite{Mateos05a}, Mateos et al.~\cite{Mateos05b}). We note that hard X-ray selected samples ($\gtrsim$ 20 keV) from the INTEGRAL and {\it Swift} satellites report marginally flatter spectral 
slopes for bright type-1 AGN, $\Gamma$$\sim$1.7-1.8 (Molina et al.~\cite{Molina09}; Winter et al.~\cite{Winter09}; Panessa et al.~\cite{Panessa08}). 
This could be explained if these surveys are biased towards sources with flatter spectral slopes or with a substantial 
Compton reflection component as these objects should be more easily detected at such high X-ray energies.

   \begin{figure*}[!ht]
   \centering
   \hbox{
   \includegraphics[angle=90,width=0.49\textwidth]{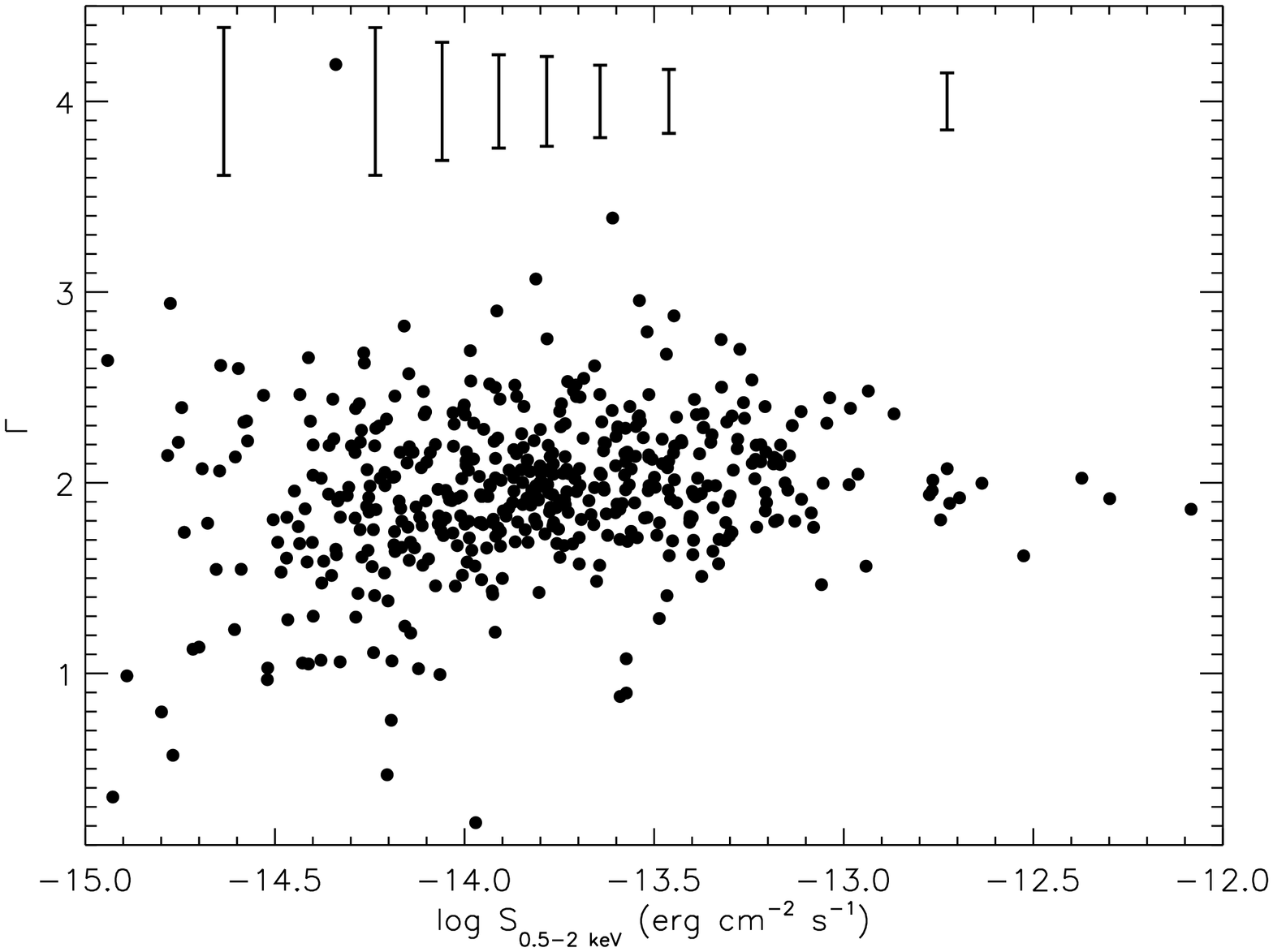}
   \includegraphics[angle=90,width=0.49\textwidth]{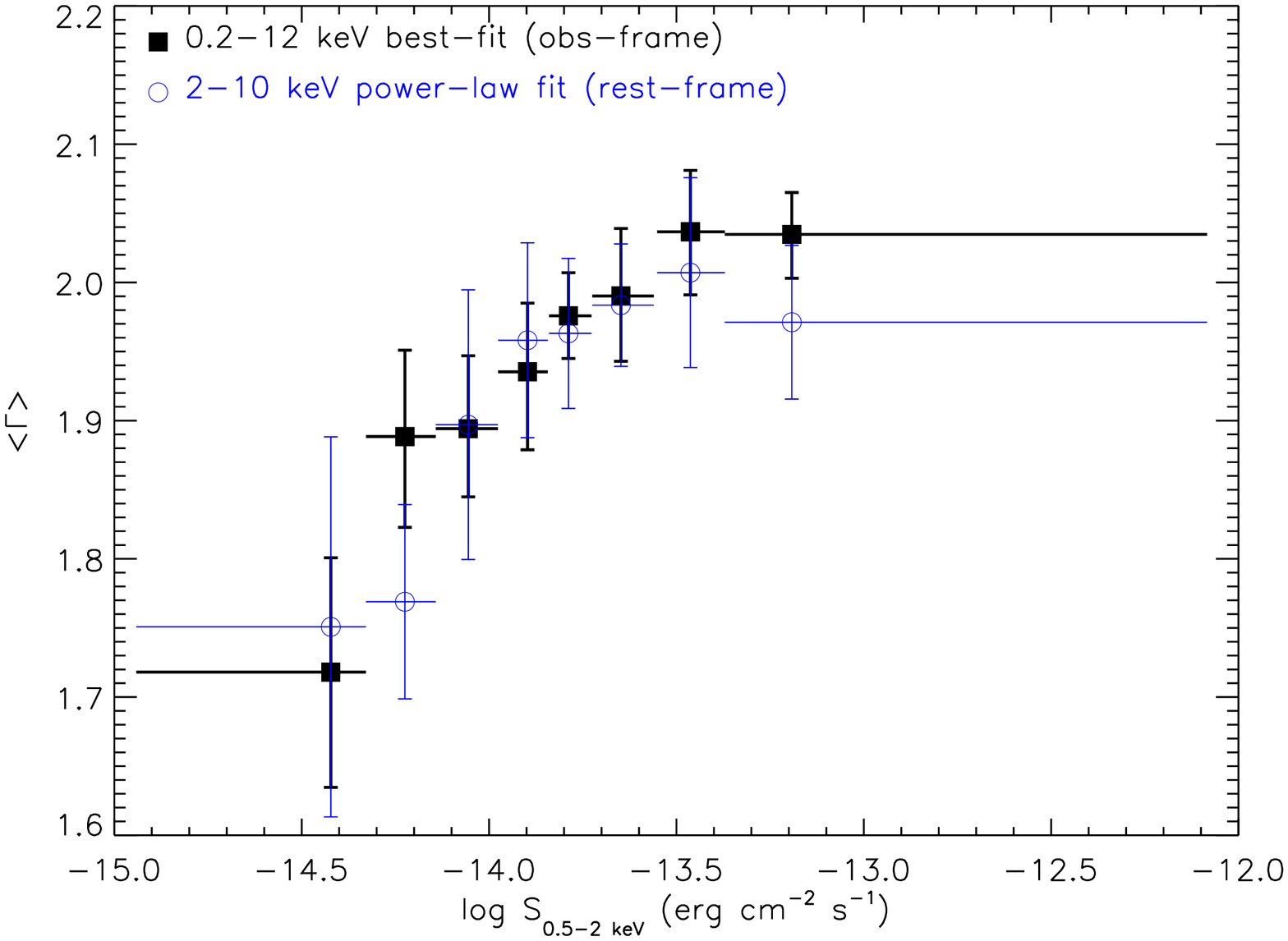}}
   \hbox{
   \includegraphics[angle=90,width=0.49\textwidth]{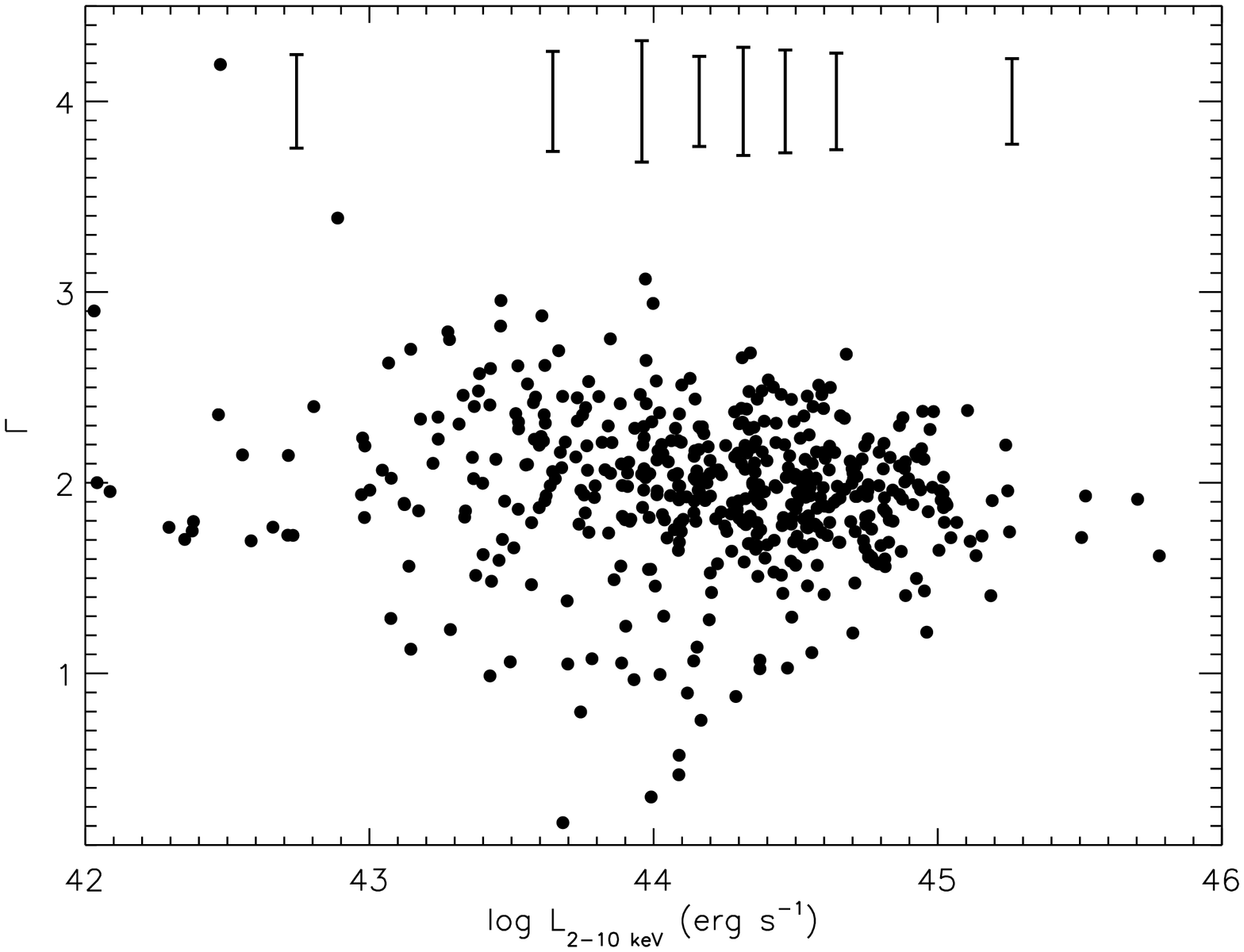}
   \includegraphics[angle=90,width=0.49\textwidth]{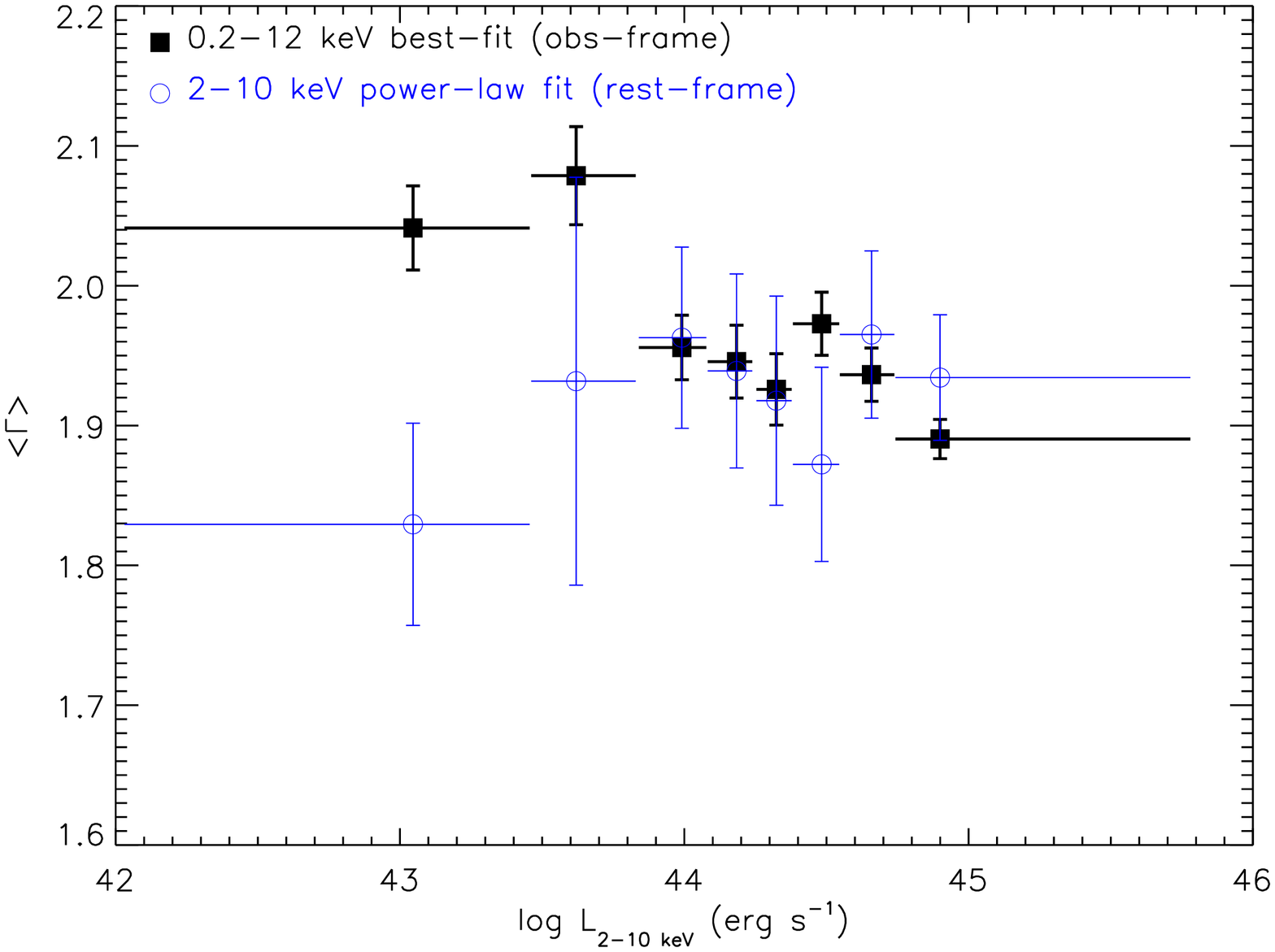}}
   \hbox{
   \includegraphics[angle=90,width=0.49\textwidth]{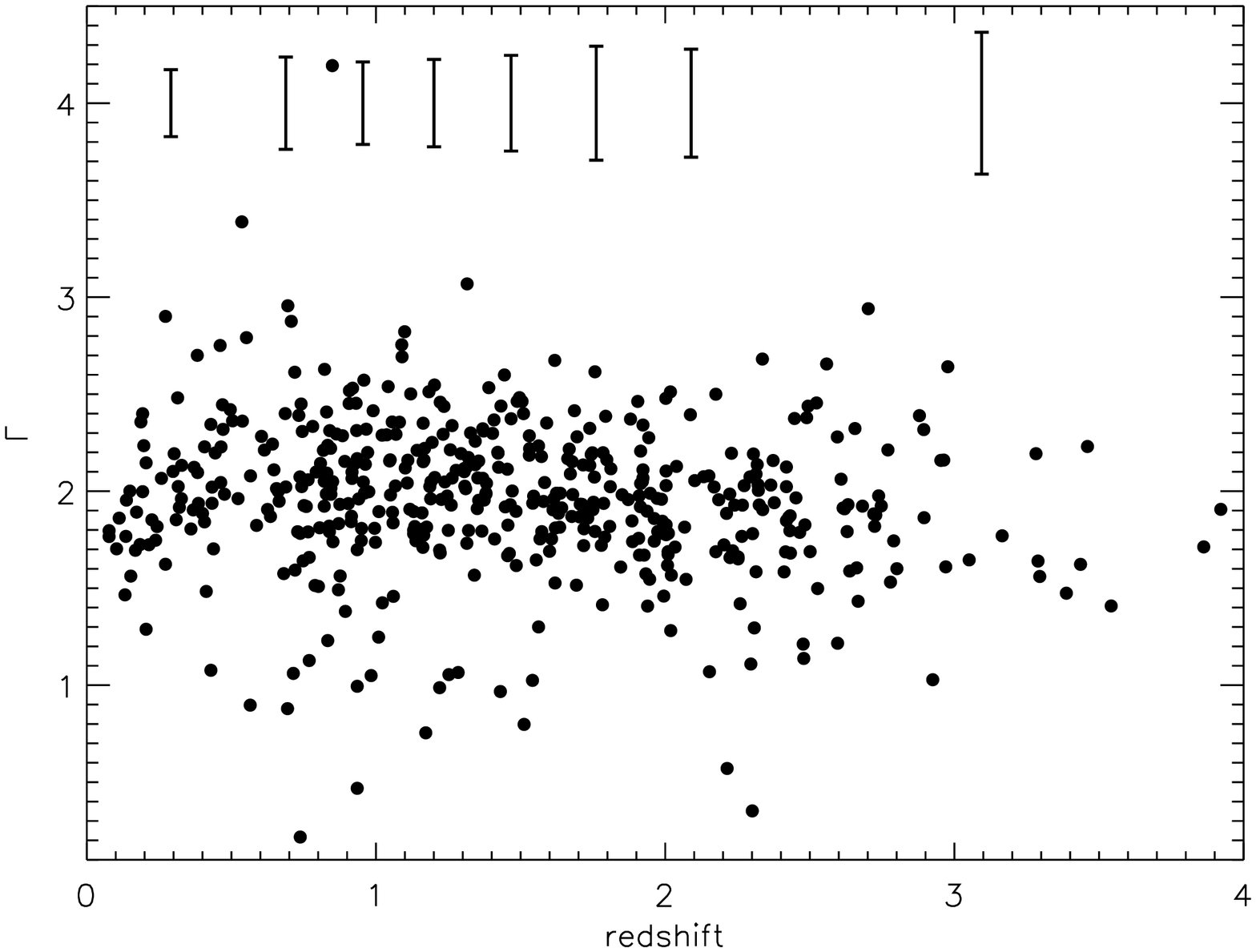}
   \includegraphics[angle=90,width=0.49\textwidth]{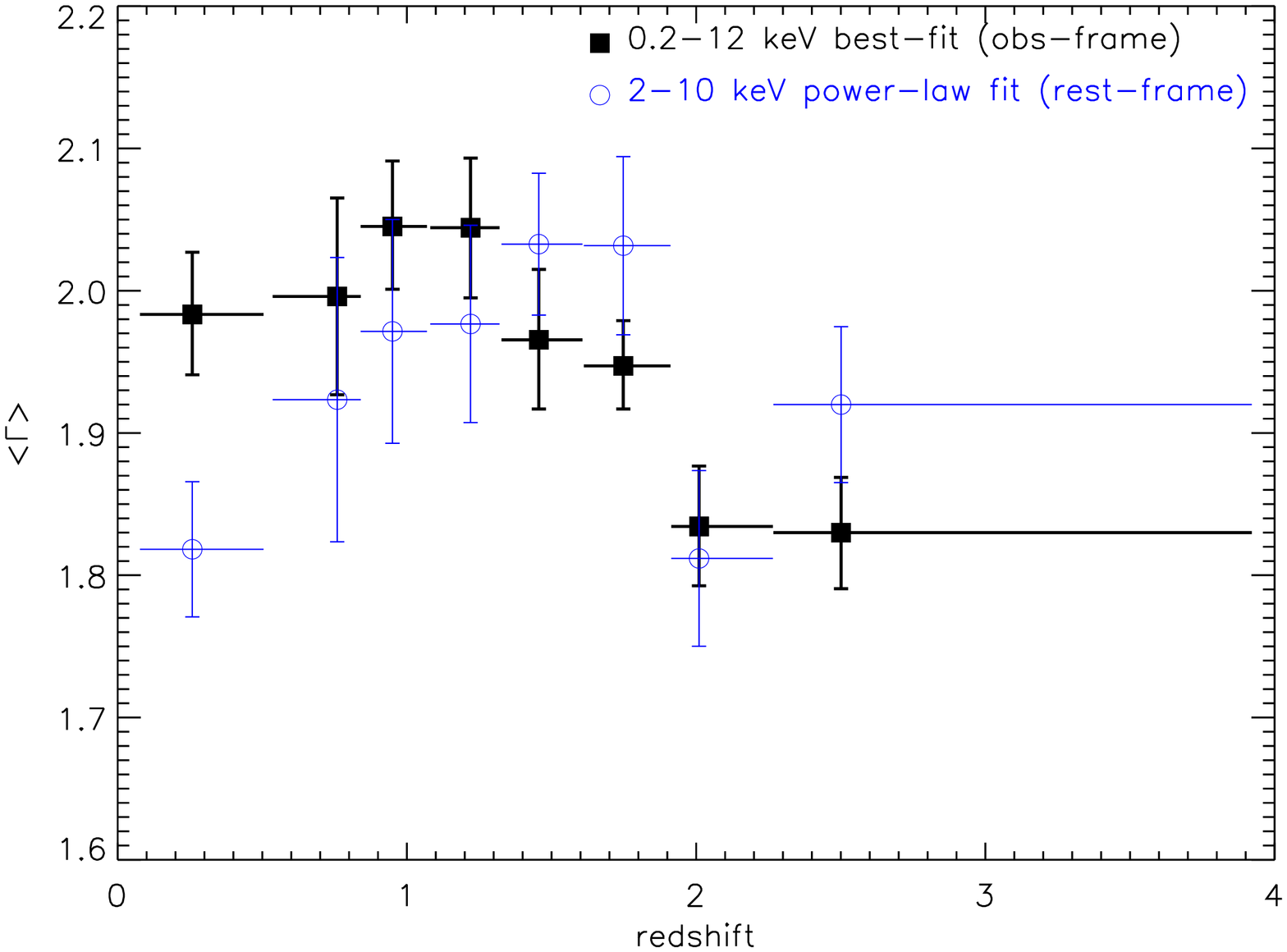}}
   \caption{Left: From top to bottom broad band continuum shape vs. 0.5-2 keV flux,
     2-10 keV luminosity and redshift respectively. The error bars at the top of the plots indicate the mean error in the measurements.
     Right: Mean spectral index of the broad band X-ray continuum, ${\rm \langle \Gamma \rangle}$, from the best-fit model 
     over the observed 0.2-12 keV band (squares) and from a power-law fitting at 2-10 keV rest-frame energies (circles) (see text 
     for details). Errors on the means are 1$\sigma$.}
              \label{Gmm_vs_flx}%
    \end{figure*}

\subsubsection{Dependence of $\Gamma$ on the flux, luminosity and redshift} 
We have investigated whether the continuum shape of type-1 AGN depends on the 
soft (0.5-2 keV) X-ray flux, hard (2-10 keV) X-ray luminosity and redshift of the sources. 
Fluxes and luminosities were computed from the best-fit model and corrected for Galactic absorption. 
Luminosities are also corrected for intrinsic X-ray absorption. 
We use hard X-ray luminosities because the measurements are less affected by uncertainties in the detected amount of 
X-ray absorption. However, to investigate the dependence of the spectral index on flux we used 
fluxes in the 0.5-2 keV band. For samples of objects derived by using different energy bands simultaneously in the source detection 
procedure such as this, the number of objects detected only in the soft band increases at faint hard (2-10 keV) fluxes. 
This is due to the fact that at faint 2-10 keV fluxes it is more difficult to detect sources with 
flat spectral slopes because their emission peaks at energies above $\sim$2 keV where the sensitivity of the X-ray detectors is much lower. 
As already stated in Mateos et al.~(\cite{Mateos05a}; see Fig. 6 in that paper) this produces an apparent softening of the objects' mean spectral 
index as we approach the survey limit in the hard band.

The left column in Fig.~\ref{Gmm_vs_flx} shows the distributions of best-fit $\Gamma$ values vs. flux, luminosity and redshift. 
For reasons of simplicity error bars in the individual measurements are not shown on the plot. The error bars at the top of the plots indicate the mean error in $\Gamma$. 
A Spearman rank correlation test indicates that $\Gamma$ correlates with the soft X-ray 
flux ($\rho$=0.18, prob=1$\times10^{-4}$), in the sense that fainter objects have flatter continuum slopes. 
On the other hand $\Gamma$ seems to be anti-correlated with both the hard X-ray luminosity ($\rho$=-0.17, prob=1$\times10^{-3}$) and 
redshift ($\rho$=-0.14, prob=2$\times10^{-3}$), i.e. $\Gamma$ becomes flatter at higher luminosities and redshifts. 
These trends become more evident when plotting $\langle \Gamma \rangle$ instead of $\Gamma$, binned in flux, luminosity and 
redshift (Fig.~\ref{Gmm_vs_flx} right column). The bins were defined to have at least 50 measurements per bin. Note the different scales used on 
the right hand plots in Fig.~\ref{Gmm_vs_flx} compared to the corresponding figures on the left. 
Mean $\Gamma$ values were obtained with the maximum likelihood method described before (squares) while the corresponding errors are the measured 
intrinsic dispersion of $\Gamma$ for 
each bin. It is unlikely that the observed trends are affected by outliers in the distribution of $\Gamma$ values in each bin, as 
these typically correspond to spectra with low S/N and hence larger associated uncertainties, and we accounted for the uncertainties in the computations 
of $\langle \Gamma \rangle$.

   \begin{figure*}[!t]
   \centering
   \hbox{
   \includegraphics[angle=90,width=1.0\textwidth]{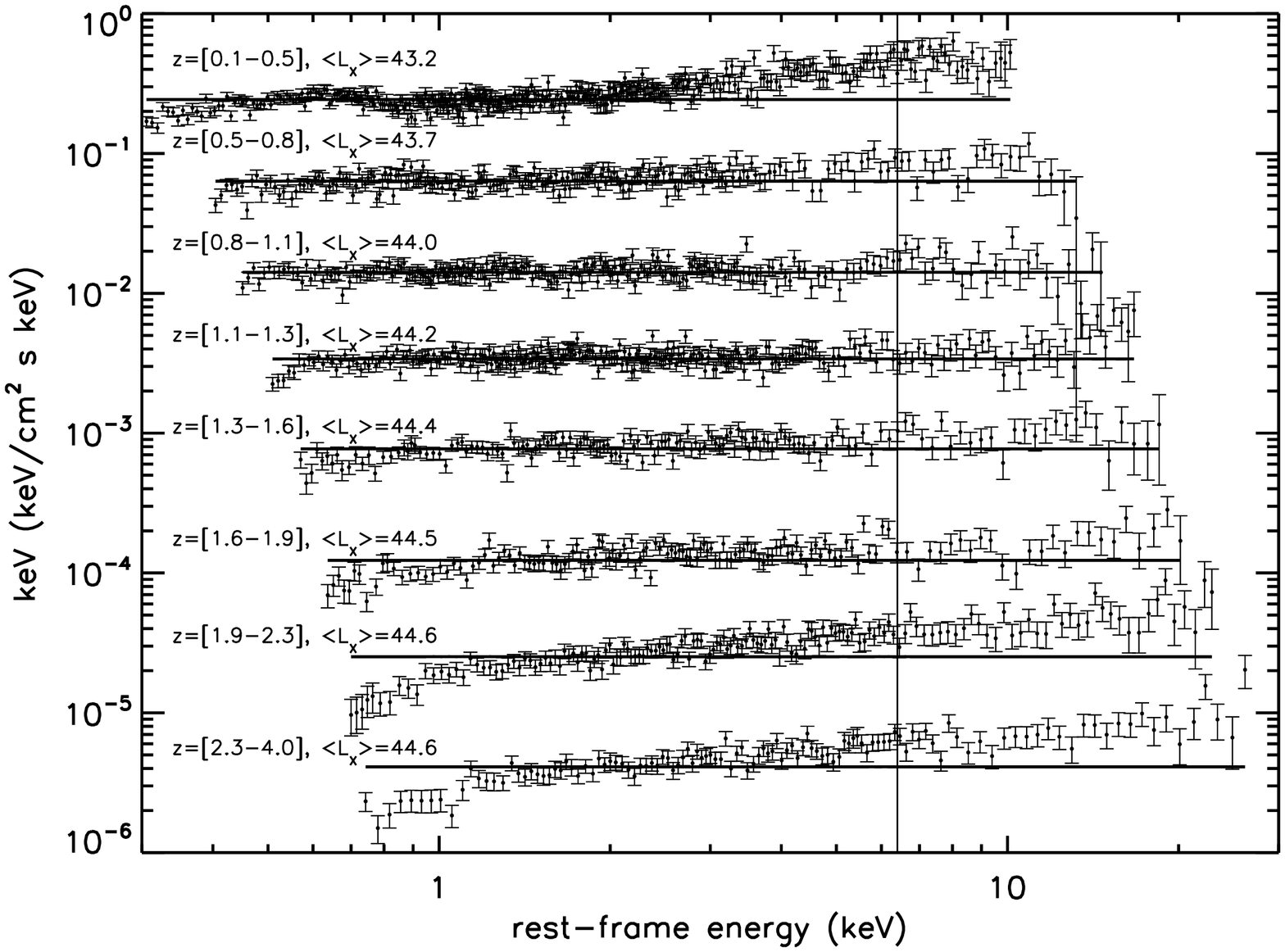}}
   \caption{Stacked rest-frame unfolded spectra of our sources in redshift bins (data points). The values at the top-left 
     of each spectra indicate the redshift range and mean 2-10 keV luminosity of the objects on each bin. 
     The horizontal lines 
     show a power-law model with photon index $\Gamma$=2 in units of $E^2 \times f(E)$ (see text for details) and the vertical line shows the 
     energy where we expect the detection of Fe K$\alpha$ line emission. The normalisations are arbitrary.}
              \label{sp_stacked}%
    \end{figure*}

We have investigated whether the spectral complexity in type-1 AGN can explain the observed trends in $\Gamma$. At rest-frame 
energies below $\sim$2 keV, intrinsic X-ray absorption and soft excess emission 
are important contributors to the X-ray spectra of AGN. Nevertheless, at low redshifts accurate constraints of $\Gamma$ become more difficult 
especially in the low count statistics regime, as the detection of soft excess or excess absorption might not be significant enough to 
be accepted by our adopted detection threshold (F-test$\ge$99\%). Furthermore, the typical absorbing column densities detected in type-1 AGN do not 
exceed a few ${\rm \times10^{22}\,cm^{-2}}$, and hence are more difficult to detect especially at high redshifts where the signatures of absorption 
shift outside the observed band-pass. Undetected spectral complexity can have an important effect on the measured values of $\Gamma$ as even 
moderate absorption (${\rm \lesssim 10^{22}\,cm^{-2}}$) can still produce apparently flatter spectral slopes than the canonical value for type-1 AGN.
In Fig.~\ref{Gmm_vs_flx} (left) we see a number of objects in our sample with best-fit $\Gamma$ values deviating more than 3$\sigma$ from the mean.
We found that all the objects with significantly flatter spectral slopes (31 objects with $\Gamma$$\lesssim$1.4) show curvature at the 
lowest energies, however the signatures of absorption are not strong enough to be statistically detected in the fitting process.
Undetected X-ray absorption can explain the apparent hardening of the mean continuum shape 
of our objects at faint soft fluxes, as the S/N of the spectra decreases with the flux as indicated  
in Fig.~\ref{Gmm_vs_flx} (top-left). On the other hand a steepening of the spectra at the softest energies, indicative of 
soft excess emission, was clearly present in the objects with the softest spectral slopes in our sample. However, in all cases the signature of this 
component was not strong enough to be detected above our significance threshold. 

In order to further investigate the scatter in the observed $\langle \Gamma \rangle$ at different fluxes, redshifts and 
luminosities, we fitted the X-ray spectra of the sources in the 2-10 keV rest-frame band a power-law model 
as the measured hard X-ray continuum of type-1 AGN is expected to be less affected by soft excess emission 
and/or absorption. We fitted only sources with at least 50 EPIC counts in the 2-10 keV rest-frame band 
($\gtrsim$60\% of the sources in the total sample). We checked that the objects with enough S/N in their 
2-10 keV rest-frame spectra have the same 0.5-2 keV flux, 2-10 keV luminosity and redshift 
distributions as the total sample. The mean continuum shapes measured in this way are shown in 
Fig.~\ref{Gmm_vs_flx} (right column, circles). In this case we have not used the maximum likelihood analysis to compute 
$\langle\Gamma\rangle$ as the typical number of 
sources per bin is only $\sim$25-30. The mean spectral slopes were obtained by weighting with the errors in the individual measurements while the error bars 
correspond to the 1$\sigma$ dispersion in the values. We found that the spectral fitting in the 2-10 keV rest-frame band returned a substantially harder $\langle \Gamma \rangle$ at the lowest luminosities. Because these objects are detected at the lowest redshifts sampled by our survey 
it is not surprising to find that the 2-10 keV rest-frame fits returned a 
substantially harder $\langle \Gamma \rangle$ at low redshifts. It is interesting 
to note that we still find sources at high redshifts ($\gtrsim$2) to be marginally flatter than those at 
lower redshifts ($\Delta\Gamma\sim0.1-0.2$) even when only rest-frame energies above 2 keV are 
used in the spectral fitting.

\subsubsection{The mean type-1 AGN spectrum vs. redshift} 
In order to understand the origin of the variation of the mean continuum shape of type-1 AGN as a function of redshift and luminosity
we stacked the spectra of the objects in each bin of redshift. The co-addition of spectra 
in such a way blurs narrow features (such as iron line emission), however this is not a problem for our analysis as 
we are interested only in measuring broad band spectral features.  
The co-added spectra were obtained following the recipe in Appendix A in Mateos et al.~(\cite{Mateos05b}). 
At high energies the S/N of the spectra decreases substantially and 
the uncertainties associated with background subtraction become much larger. Hence we only stacked spectra from 0.5 keV (to minimise EPIC calibration 
uncertainties at the lowest energies) to 8 keV (observed frame). The stacked spectra were shifted to rest-frame energies using the mean redshift of each bin. 
Fig.~\ref{sp_stacked} shows both the unfolded stacked spectra (data points) and 
a power-law model \footnote{$f(E)=dN/dE=A \times E^{-\Gamma}$ where {\it N} is the number of photons, {\it E} the photon 
energy, {\it A} the normalisation and $\Gamma$ is the photon index.} in units of $E^2\times f(E)$ (horizontal solid lines) with 
photon index $\Gamma$=2 ($\sim$mean spectral index of the broad band X-ray continuum of our objects). 
The vertical line shows the energy where we expect the iron fluorescence (Fe K$\alpha$) line emission. 
Also indicated are the redshift range and mean 2-10 keV luminosity of each bin. The 
normalisations are arbitrary. We only show for clarity pn stacked spectra, however MOS and pn results 
are consistent. We checked that the stacked spectra are not dominated by bright objects in the bins. 

\begin{table}[!t]
  \caption{Summary of the results of the spectral fitting of the type-1 AGN stacked spectra.}
\label{table:2}      
\centering                          
\begin{tabular}{c c c c c c c c c c c c c c c c c c c}        
\hline\hline                 
${\rm \langle z \rangle}$ &  ${\rm \Gamma_{best-fit}}$ & ${\rm {\it kT}_{best-fit}}$ & ${\rm \Gamma_{2-10\,keV}}$\\
(1) & (2) & (3) & (4) \\
\hline                        
 0.31 & 1.69$\pm$0.03 & 0.11$\pm$0.01 & 1.60$\pm$0.07\\
 0.74 & 2.02$\pm$0.04 & 0.09$\pm$0.01 & 1.83$\pm$0.10\\
 0.94 & 1.97$\pm$0.03 &    -          & 1.96$\pm$0.08\\
 1.19 & 2.01$\pm$0.03 &   -           & 1.93$\pm$0.04\\
 1.45 & 1.93$\pm$0.06 &   -           & 2.01$\pm$0.17\\
 1.74 & 1.99$\pm$0.05 &   -           & 1.91$\pm$0.11\\
 2.01 & 1.81$\pm$0.05 &   -           & 1.81$\pm$0.12\\
 2.56 & 1.81$\pm$0.05 &   -           & 1.77$\pm$0.20\\
\hline                                   
\end{tabular}

(1) Mean redshift in bin. (2) Best-fit $\Gamma$ value.
(3) Rest-frame temperature of the blackbody model (in keV) used to fit the soft excess component.
(4) Best-fit $\Gamma$ from a power-law fitting in the 2-10 keV rest-frame band.
Errors are 90\% confidence.
\end{table}

The results of fitting the co-added spectra are summarised in Table~\ref{table:2}. 
The spectra of objects at redshifts above $\sim$0.9 were best fitted with a simple power-law model. 
Cold absorption was required in all spectra, however the measured values were consistent with typical 
Galactic values (${\rm \lesssim10^{21}\,cm^{-2}}$).
At lower redshifts a soft excess component was detected with an F-test significance $>$99.99\%. 
We best modelled this component with a thermal blackbody. The measured temperatures are listed in Table~\ref{table:2}. 
For comparison we also show in Table~\ref{table:2} the best-fit spectral slopes obtained by fitting 
the spectra with a power-law in the 2-10 keV rest-frame band. 
The results of this analysis show the same dependence of $\langle \Gamma \rangle$ on the redshift as found 
in Fig.~\ref{Gmm_vs_flx}, i.e. the mean continuum shape of our objects becomes substantially 
harder ($\Delta\Gamma$$\sim$0.2-0.3) at the lowest and highest redshifts sampled by our survey. 

An apparently harder spectral slope could be ascribed to the net effect of stacking spectra with different 
amounts of X-ray absorption and marginally different redshifts. Another possible explanation for the hard spectral slope could be a hardening of 
the continuum at the highest energies due to the presence of the Compton reflection bump. However, reflection is unlikely to explain the hard $\Gamma$ at the lowest redshifts 
as the reflection component only becomes important at rest-frame energies above $\sim$10 keV, which are not sampled in our lowest redshift 
bin. Furthermore a power-law fit to the stacked spectrum at rest-frame energies above 4 keV (where no effect from X-ray absorption is expected) resulted 
in a steeper spectral slope, $\Gamma$=1.9$\pm$0.2. Therefore our results indicate that the overall harder spectral slope at the 
lowest redshifts is an effect of X-ray absorption.

At high redshifts absorption is unlikely to explain the measured harder continuum of the objects, as at redshifts $\gtrsim$2 
we expect most of the spectral signatures of moderate 
absorption to be shifted outside the observed band-pass. The flattening of $\Gamma$ could be the result of the redshifting 
of the Compton reflection bump into the observed band and/or the redshifting of the soft excess outside the observed band.
However, no strong indications for spectral evolution of type-1 AGN have been detected so far, as the power-law slopes of high 
redshift QSOs (out to $z\gtrsim$4) have been constrained to a value $\Gamma$$\sim$2 
(Just et al.~\cite{Just07}; Shemmer et al.~\cite{Shemmer06}; Vignali et al.~\cite{Vignali05}). 
It is therefore unlikely that the Compton reflection bump is producing the overall flatter spectral slopes in our sources. 

\subsubsection{Effect of the object detection efficiency on the measured $\langle\Gamma\rangle$} 
We have investigated whether the flatter spectral slope for high redshift sources can be explained by the dependence of the efficiency of detection of the XMM-{\it Newton} EPIC cameras on the spectral shape of the objects: in high redshift objects we are sampling 
rest-frame energies above 10 keV and hence harder sources might be more easily detected than steep spectrum sources or sources with no reflection component.
This is related to the fact that source detection with X-ray telescopes is in photons rather than in flux, hence different flux limits are 
sampled for sources with different intrinsic spectra (Carrera et al.~\cite{Carrera07}; Della Ceca et al.~\cite{Ceca99}; Zamorani et al.~\cite{Zamorani88}). 

In order to constrain the importance of this effect we have simulated power-law spectra with different continuum shapes at different redshifts. 
We have performed the simulations using response files for the EPIC pn camera only, although the results of this analysis also apply to the EPIC MOS 
cameras. We required the simulated spectra to have a 0.2-12 keV flux of ${\rm 10^{-14}\,erg\,cm^{-2}\,s^{-1}}$, typical for the objects 
in our sample. At each redshift we found the luminosity that corresponds to the selected flux for a typical spectral slope $\Gamma$=1.9. Then we used that luminosity 
to obtain the normalisation of the simulated spectra as a function of the spectral index. We then computed the number of counts in the 
0.2-12 keV band. 

\begin{figure}[!t]
   \centering
   \hbox{
   \includegraphics[angle=90,width=0.5\textwidth]{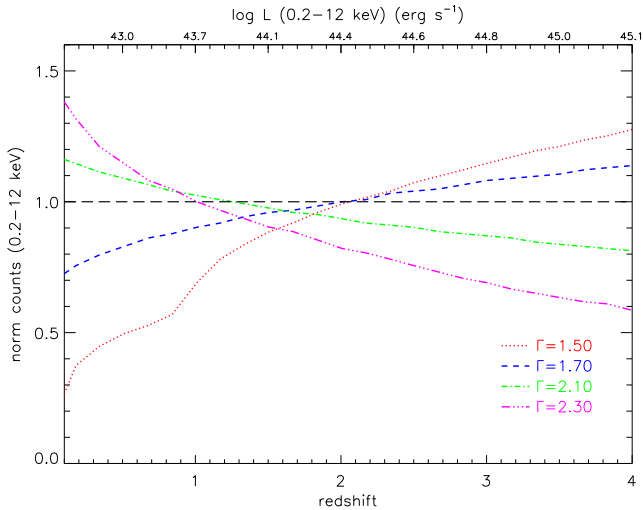}}
   \caption{Efficiency of detection of objects with the EPIC-pn camera as a function of redshift and spectral shape.
     We defined the efficiency as the detected number of counts in the 0.2-12 keV band normalised to the number of 
     counts detected for a power-law spectrum with $\Gamma$=1.9 (horizontal line).
     The top axis shows the luminosities that are required to obtain a flux of ${\rm 10^{-14}\,erg\,cm^{-2}\,s^{-1}}$ at different redshifts for a power-law spectrum 
     with $\Gamma$=1.9.}
              \label{sims_gamma}%
    \end{figure}

The change in detection efficiency as a function of redshift and spectral slope is shown in Fig.~\ref{sims_gamma}, where the y-axis represents 
the detected number of counts in the broad 0.2-12 keV band normalised to the number of counts detected for a power-law spectrum with $\Gamma$=1.9. 
The top axis shows the luminosities that were used to normalise the spectra at different redshifts. It is evident that there is a strong dependence 
of the sensitivity of detection of objects on the spectral shape and this also has a strong dependence on redshift. As we move to higher redshifts the efficiency 
of detection of soft objects decreases substantially while the efficiency of detection of hard objects increases. 
Therefore in flux limited samples such as this, the strong dependence of the efficiency of detection on the spectral shape at different redshifts 
could produce the apparent hardening of the mean continuum shape of objects detected at high redshifts: as we move to higher redshifts, at the flux limit of the survey 
we better detect objects in the 'hard tail' of the distribution of continuum spectral slopes. 
This effect was already noted by Francis et al.~(\cite{Francis93}), who pointed out that the range of continuum slopes has an important 
effect upon the measured spectral evolution of the AGN population. Because of this important effect we cannot confirm whether the 
intrinsic broad band continuum of type-1 AGN has any dependence on redshift (and hence on the hard X-ray luminosity) without doing 
extensive simulations, which are beyond the scope of this paper.

There have been claims in the literature of an anti-correlation between the spectral slope and 
the hard X-ray luminosity, in the sense that the X-ray spectral slope hardens at higher luminosities 
(e.g. Young et al.~\cite{Young09}; Saez et al.~\cite{Saez08}; Page et al.~\cite{Page05}). However, other studies 
find the opposite correlation, i.e a softening of the spectral slope at higher luminosities (Dai et al.~\cite{Dai04}) or 
no correlation at all (e.g. George et al.~\cite{George00}; Reeves \& Turner~\cite{Reeves00}; Winter et al.~\cite{Winter09}). The detected hardening of the spectral slope 
at higher hard luminosities in our sample is dominated by the two lowest luminosity bins where 
the uncertainties in the measurements of $\Gamma$ are larger due to spectral complexity. 
Indeed, at luminosities above ${\rm \sim10^{44}\,erg\,s^{-1}}$ $\langle \Gamma \rangle$ seems to remain fairly constant over $\sim$two orders of magnitude in 
luminosity, so we do not have strong indications of a dependence of the spectral slope with the 2-10 keV luminosity. We note also that the strong dependence of the 
sensitivity of detection of sources on the spectral shape must have an impact on the measured mean continuum shapes at different luminosities in flux 
limited samples due to the strong correlation of the X-ray luminosity and redshift in these surveys.

\subsection{X-ray absorption}
We did not detect absorption in excess of the Galactic value in most of our type-1 AGN.
Only the spectra of 19 objects out of 487 showed strong enough signatures of absorption to be detected above our significance 
threshold. This value corresponds to 3\% of the objects 
after taking into account the spurious detections for the selected F-test significance.  
Furthermore, a number of objects in our sample have best-fit $\Gamma$ values deviating more than 3$\sigma$ from the mean (see Fig.~\ref{Gmm_vs_flx}).
Using a fixed spectral slope of $\Gamma$=1.9 we detected substantial X-ray absorption in all sources with typical measured 
column densities ${\rm N_H\sim a\,few\times 10^{22}\,cm^{-2}}$. Previous studies of type-1 AGN selected in the 
optical (Piconcelli et al.~\cite{Piconcelli05}) and X-ray band (Mainieri et al.~\cite{Mainieri07}; Page et al.~\cite{Page06}; 
Mateos et al.~\cite{Mateos05b}; Mateos et al.~\cite{Mateos05a}; Caccianiga et al.~\cite{Caccianiga04}; Perola et al.~\cite{Perola04}; Barcons et al.~\cite{Barcons02}) 
report a fraction of absorbed type-1 AGN of 10\% or less, i.e marginally higher than our reported value, however most of these studies used lower confidence thresholds 
when applying the F-test. 
However, accounting for all sources with measured flat spectral slopes, the fraction of X-ray absorbed objects in our sample increases to $\approx$8\%, a 
value entirely consistent with previous measurements.

   \begin{figure}
   \centering
   \hbox{
   \includegraphics[angle=90,width=0.49\textwidth]{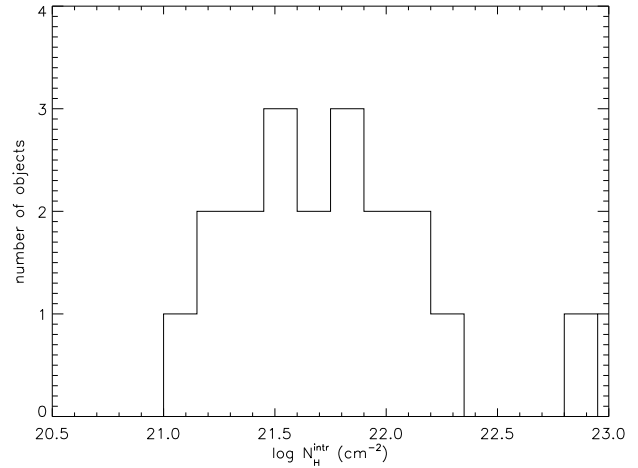}}
   \caption{Distribution of rest-frame absorbing column densities.
   }
              \label{nh_dist}%
    \end{figure}

   \begin{figure}[!th]
   \centering
   \hbox{
   \includegraphics[angle=90,width=0.49\textwidth]{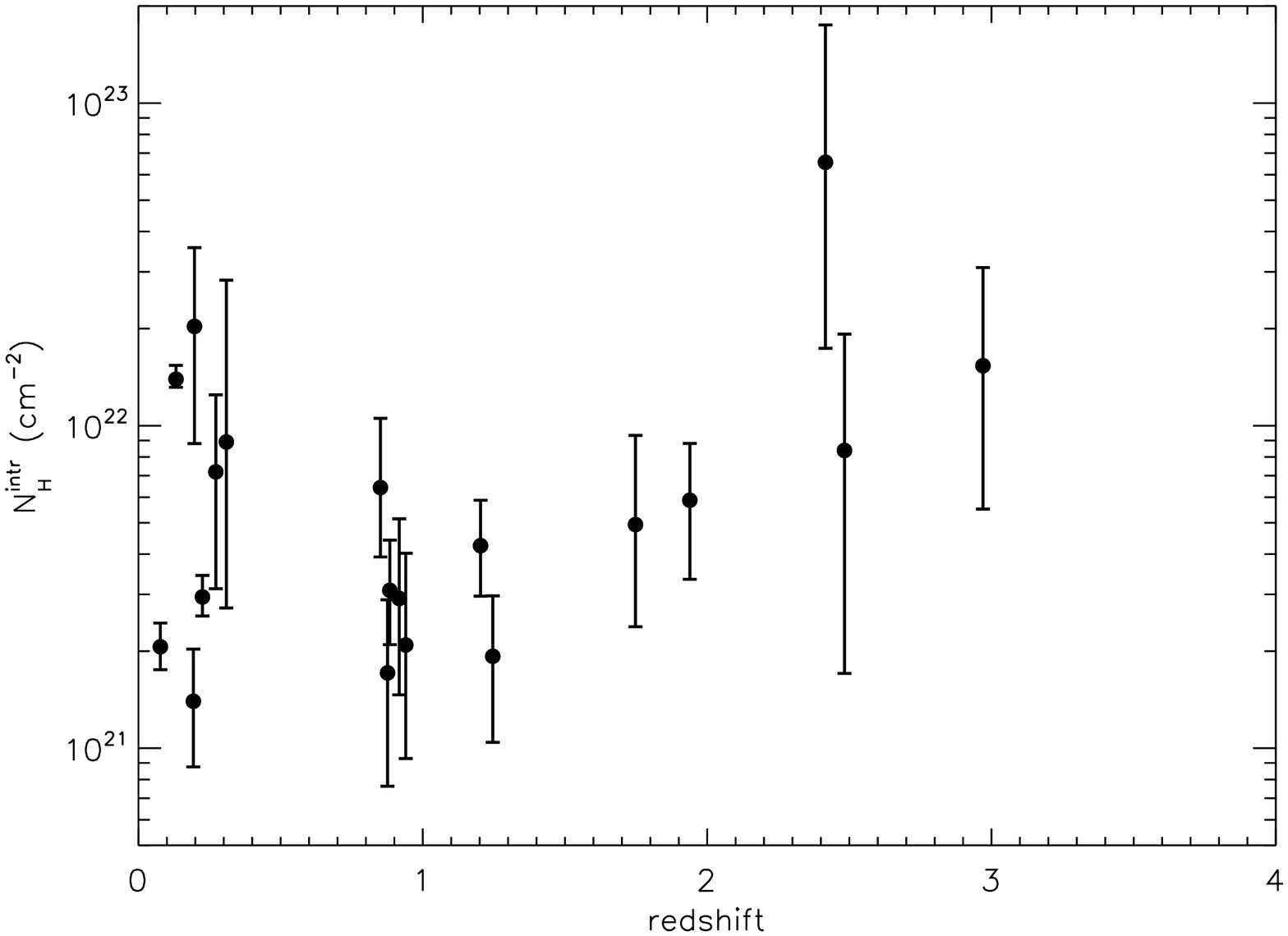}}
   \hbox{
   \includegraphics[angle=90,width=0.49\textwidth]{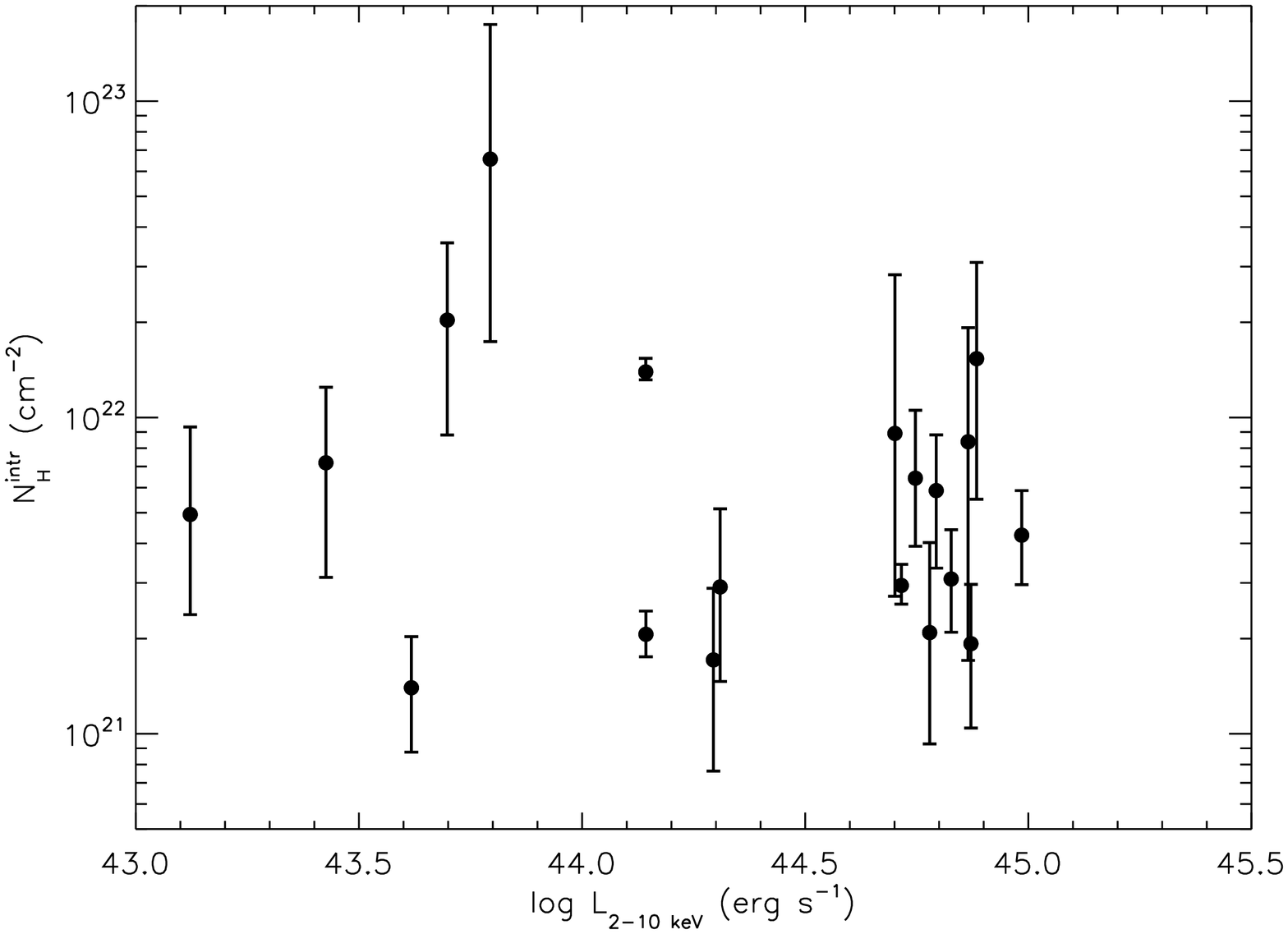}}
   \caption{X-ray absorption vs. redshift (top) and hard (2-10 keV) X-ray luminosity (bottom). Errors are 90\% confidence.
     }
              \label{nh_vs_lum}%
    \end{figure}

The distribution of absorbing column densities for our type-1 AGN is shown in Fig.~\ref{nh_dist}. The measured amounts of X-ray absorption 
are ${\rm \sim few\times10^{22}\,cm^{-2}}$, i.e. lower than the amount of absorption typically detected for type-2 AGN 
(Krumpe et al.~\cite{Krumpe08}; Mateos et al.~\cite{Mateos05b}). Only one object shows an absorbing column density substantially 
above ${\rm 10^{22}\,cm^{-2}}$, but with very large uncertainty. 
Fig.~\ref{nh_vs_lum} shows the distribution of X-ray absorption as a function of redshift and 2-10 keV luminosity. We do not 
see any clear correlation amongst these parameters. We only find a marginal indication that the minimum detected 
column density increases with redshift. However, this effect can easily be explained in terms of 
absorption being redshifted towards lower energies and further decreased by a factor $\sim$$(1+z)^{2.7}$ (Barger et al.~\cite{Barger02}).

We found little or no evidence for X-ray absorption in the great majority of our type-1 AGN, in line with AGN unification models. However $\gtrsim$3\% of type-1 AGN show substantial absorption in X-rays. The apparent mismatch 
between the optical classification and the X-ray properties of these objects is a challenge for 
the simplest AGN unification model where the type-1/type-2 AGN classification is due solely to an orientation effect.

Dedicated monitoring of a few bright AGN has found strong variations in X-ray absorption on time 
scales as short as a few hours (Risaliti et al.~\cite{Risaliti02}). These results 
put tight constraints on the location of the X-ray absorber, suggesting that the toroidal obscuration region used in the AGN unification 
schemes to explain the observed AGN diversity might be co-spatial with the 
broad line region. In the clumpy torus models recently presented by Nenkova et al.~(\cite{Nenkova08}) there is a physical reason that 
could explain the detection of absorption in a fraction of type-1 AGN: the broad line region clouds located inside the dust sublimation radius 
are dust free, and hence affect X-rays but not the optical. In this scenario optical extinction should always 
be lower that the expected value assuming the Galactic dust-to-gas ratio as it is typically observed (Maiolino et al.~\cite{Maiolino01a}). 
On the other hand, it has also been suggested that if the dust composition in the circumnuclear region of AGN has different properties 
than in the Galactic diffuse interstellar medium, for example if it is dominated by large grains, this medium will be less effective in absorbing 
the optical/UV radiation (Maiolino et al.~\cite{Maiolino01b}).

A large fraction ($\sim$50\%) of both Seyfert 
1s and QSOs show signatures of highly ionised gas along the line of sight, the 
so-called 'warm absorber', in their X-ray spectra (e.g. Piconcelli et al.~\cite{Piconcelli05}; George et al.~\cite{George98}).
These absorbers, probably associated with outflowing winds, have been found to show a wide range of ionisation, have typical outflowing 
velocities of a few hundred km s$^{-1}$ and have column densities in the range ${\rm 10^{22-23}\,cm^{-2}}$. Warm absorption could also explain the apparent mismatch 
between the optical and X-ray classification in our absorbed type-1 AGN. 

\subsection{Soft excess} 
A strong broad excess emission at energies below $\sim$2 keV with respect to the  
extrapolation of the 2-10 keV continuum, known as {\it soft excess}, 
is known to be an important component of the X-ray spectra of Seyfert 1 galaxies and 
radio-quiet quasars (e.g. Porquet et al.~\cite{Porquet04}).
The nature of the soft excess emission in unobscured AGN is still one of the open questions in our understanding of the 
emission properties of AGN. 

   \begin{figure}[!t]
   \centering
   \hbox{
   \includegraphics[angle=90,width=0.49\textwidth]{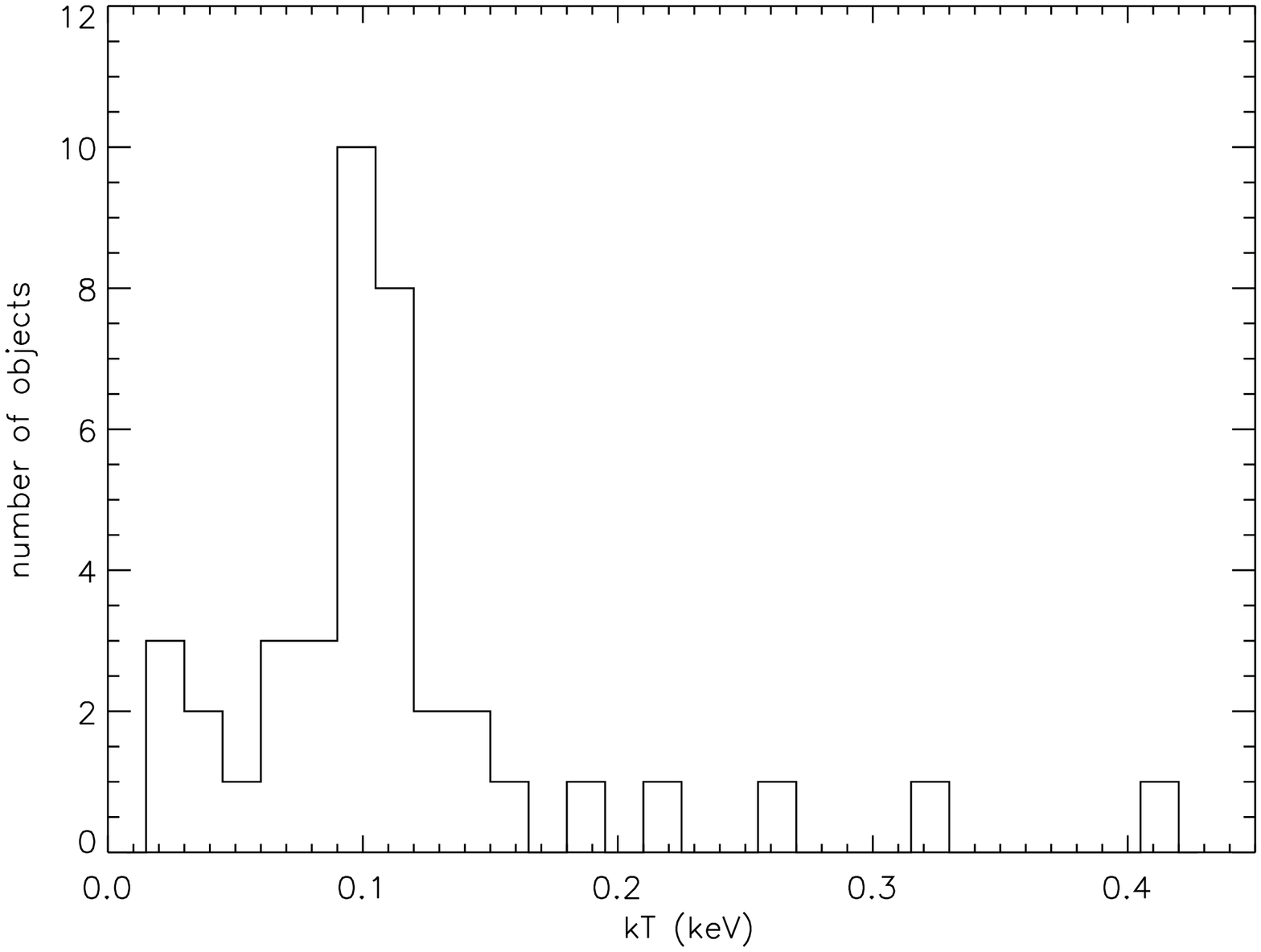}}
   \hbox{
   \includegraphics[angle=90,width=0.49\textwidth]{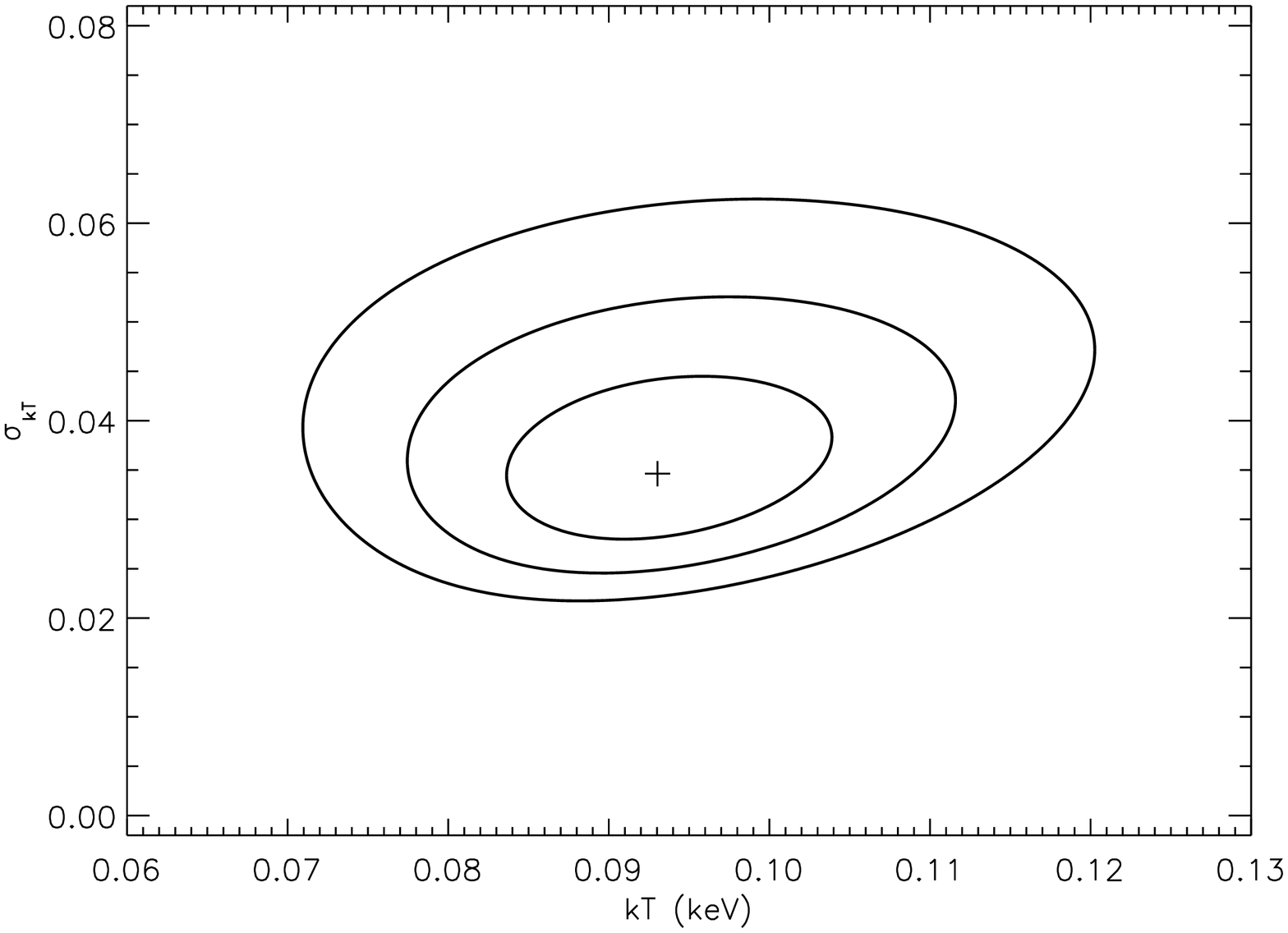}}
   \caption{Top: Distribution of rest-frame temperatures of the soft excess component when 
     modelled with a thermal blackbody. Bottom: Best-fit values of $kT$
     and its intrinsic dispersion $\sigma_{kT}$ from the maximum likelihood 
     analysis (cross) as well as the corresponding 1$\sigma$, 2$\sigma$ and 3$\sigma$ 
     confidence contour levels.
   }              
   \label{kT_dist}%
    \end{figure}

In order to search for this spectral component we have used a 
thermal blackbody model that, to a first order approximation, can model the thermal emission from an optically-thick accretion disk. 
We also fitted soft excess signatures with a partial covering model but in all cases we obtained better quality fits with the blackbody 
model. Soft excess has been detected in 40 objects with an F-test significance $\ge$99\%. Considering only objects at redshifts 
below 0.5, where most soft excess emission is detected, we found that the fraction of objects with detected soft excess increases to $\sim$36\%. A soft 
excess emission component seems to be very common in type-1 AGN. For example, studies based on ASCA (George et al.~\cite{George00}; 
Reeves \& Turner~\cite{Reeves00}) reported the detection of soft excess emission in $\sim$50-60\% of the 
objects while in a more recent study of a sample of QSOs from the Palomar-Green (PG) Bright Quasar Survey this spectral component was detected in $\sim$90\% of the 
objects (Piconcelli et al.~\cite{Piconcelli05}).

   \begin{figure*}[!ht]
   \centering
   \hbox{
   \includegraphics[angle=90,width=0.49\textwidth]{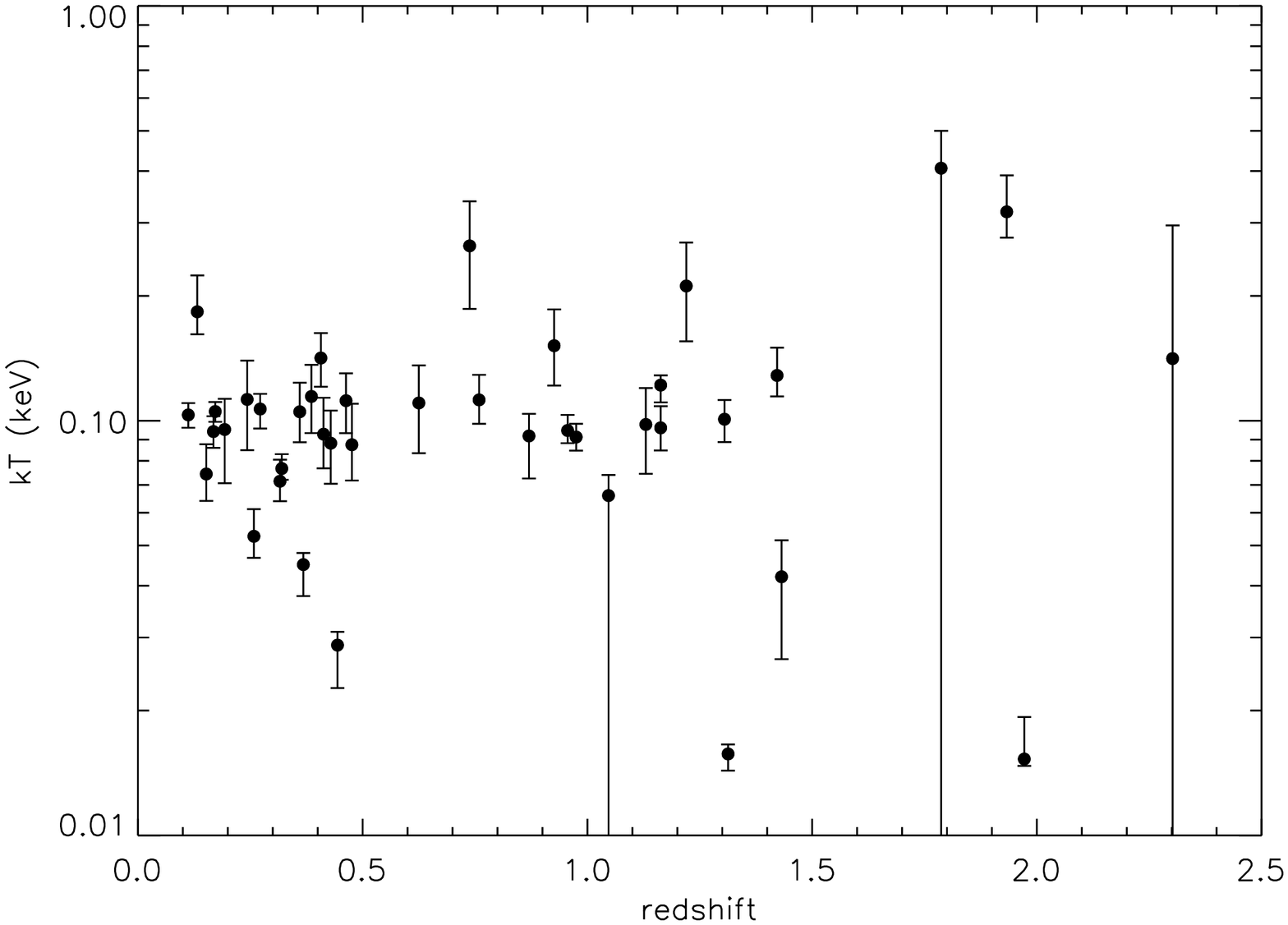}
   \includegraphics[angle=90,width=0.49\textwidth]{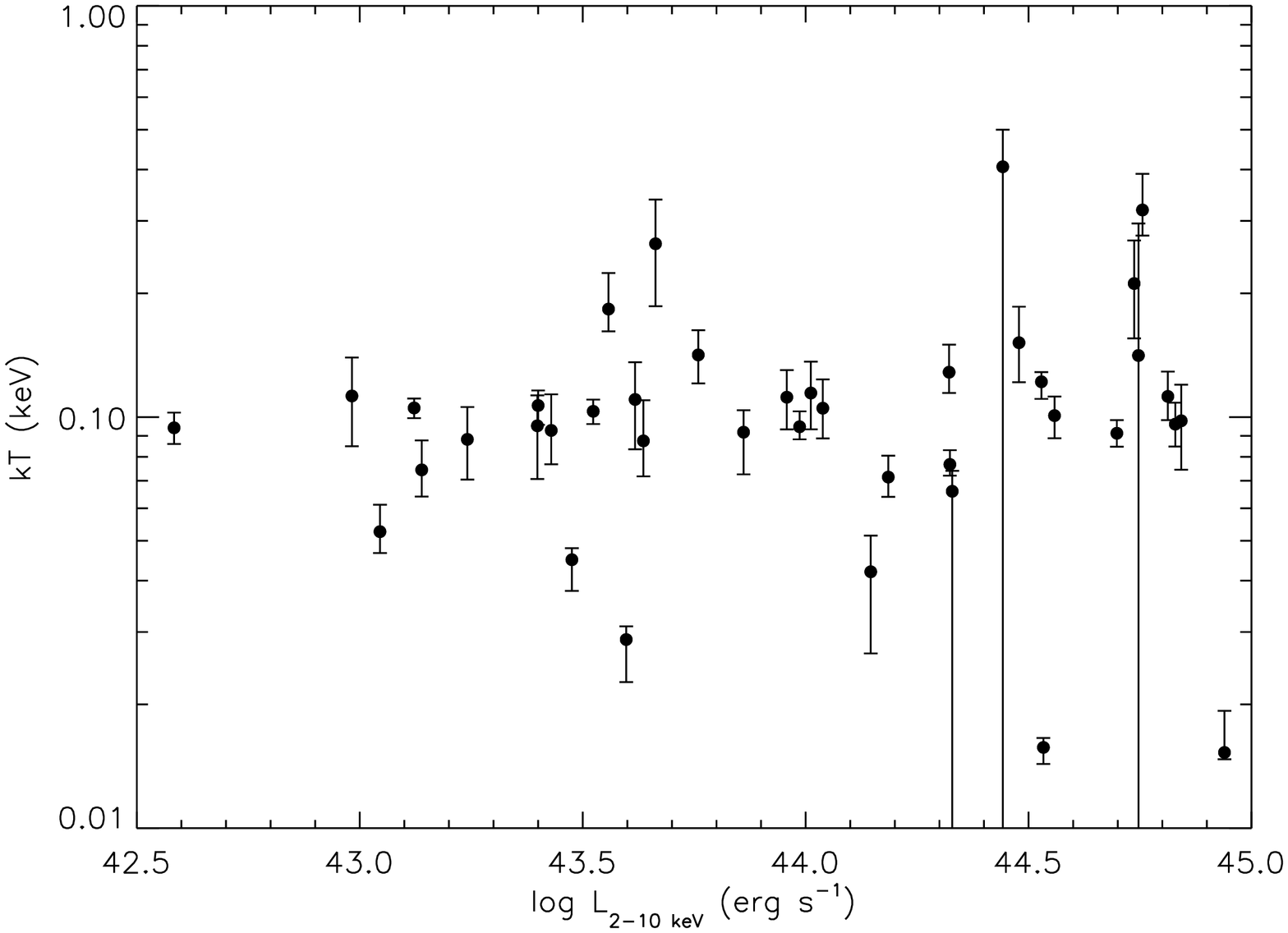}}
   \hbox{
   \includegraphics[angle=90,width=0.49\textwidth]{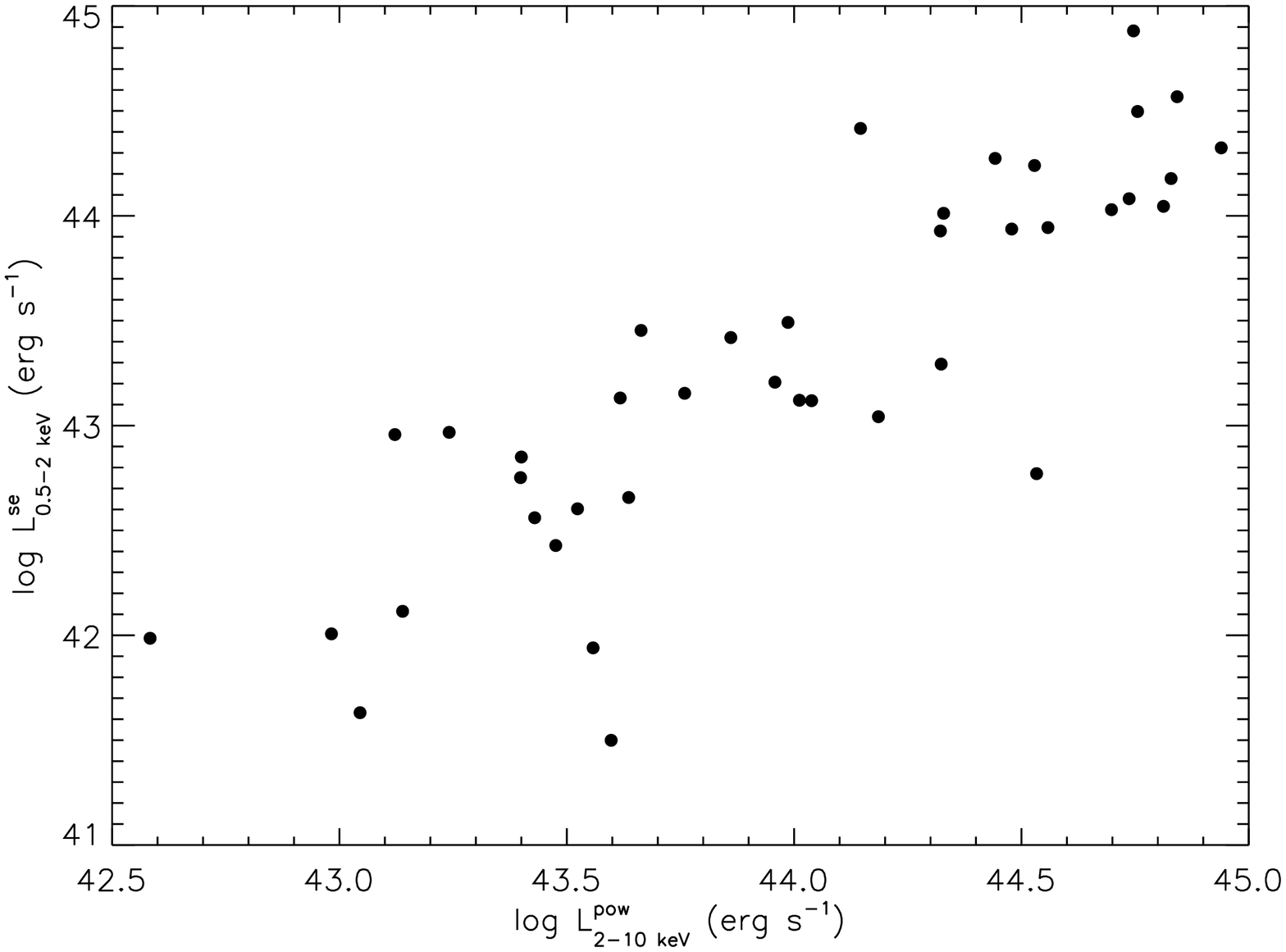}
   \includegraphics[angle=90,width=0.49\textwidth]{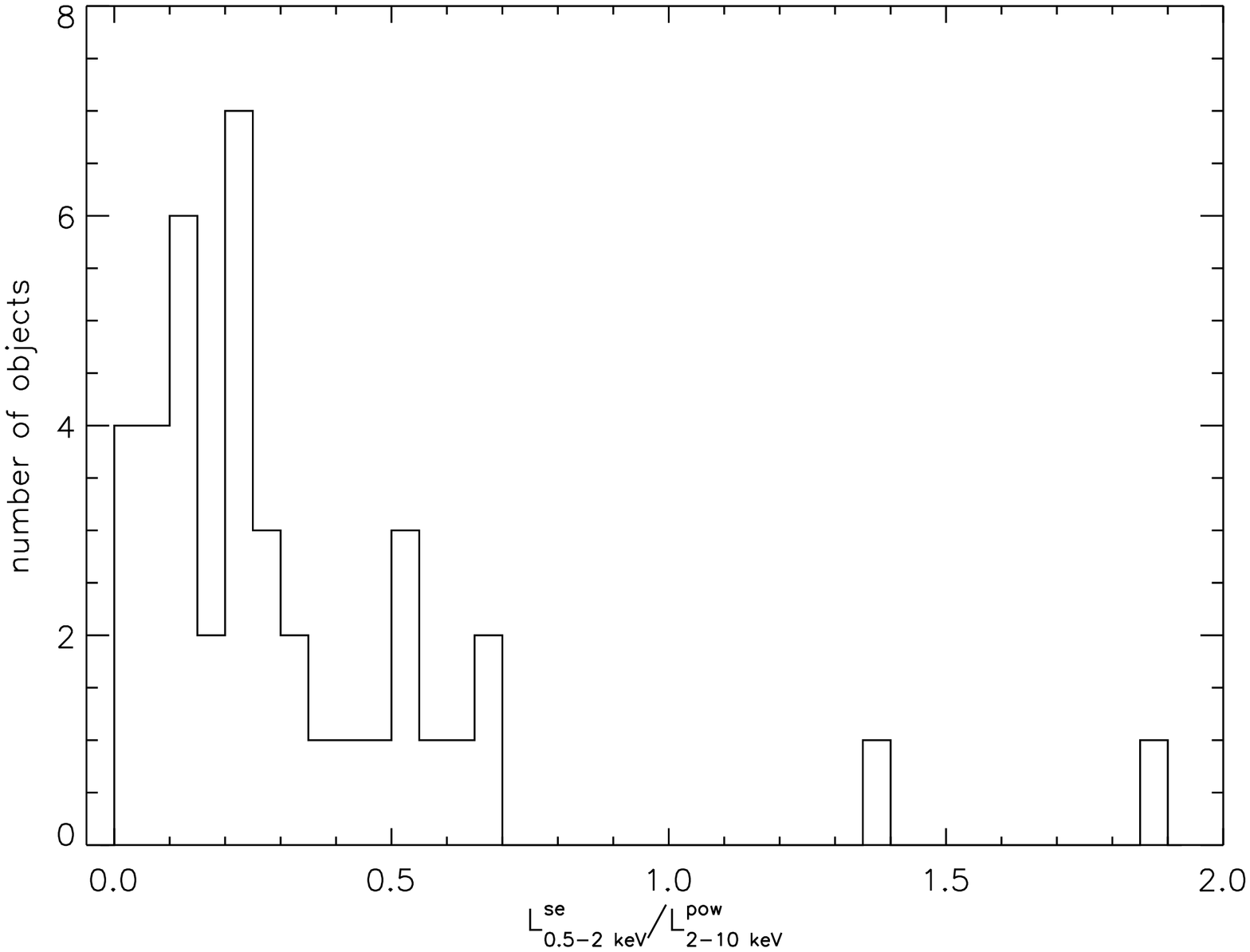}}
   \caption{Top row: Temperature of the blackbody model used to fit the soft excess vs. redshift (left) and hard (2-10 keV) luminosity (right).
     Bottom row: Luminosity of the soft excess vs. luminosity of the intrinsic power-law continuum (left).
     Distribution of the strength of the soft excess measured as the ratio between the soft excess and power-law luminosities (right). 
   }              
   \label{kT_prop}%
    \end{figure*}

Taking into account the fraction of spurious detections we detected 
soft excess in 7.3\%$\pm$2\% of the full sample of XWAS type-1 AGN at all redshifts. That this value is so much lower that 36\% is quite plausibly a consequence of the 
soft excess being redshifted below the EPIC energy range for a large fraction of our XWAS spectra, and hence 7.3\% can be considered 
as a very conservative lower limit to the fraction of type-1 AGN with soft excess emission.

We searched for warm absorbers in our objects with detected soft excess emission fitting the spectra 
with the warm absorber model {\it absori} in XSPEC. In all cases the warm absorber did not provide a good quality fit to the data. 
Therefore here we describe the properties of the soft excess emission fitted with the thermal blackbody model.

Fig.~\ref{kT_dist} (top) shows the distribution of measured temperatures of the soft excess component. We clearly see that the 
values are narrowly distributed and they tend to cluster around {\it kT}$\sim$0.1 keV in the rest-frame. 
We constrained the mean value of the temperature of the soft excess and the amplitude of the 
intrinsic scatter assuming that the distribution of temperatures can be reproduced with a Gaussian of mean ${\rm \langle {\it kT} \rangle}$ and 
dispersion ${\rm \sigma_{\langle {\it kT} \rangle}}$. The best simultaneous estimates were obtained with a maximum likelihood 
technique accounting for both the errors in the measurements and the intrinsic dispersion 
of values (Maccacaro et al.~\cite{Maccacaro88}).
Fig.~\ref{kT_dist} (bottom) shows the 1$\sigma$, 2$\sigma$ and 3$\sigma$
contours for the two parameters together with the best-fit values (cross).
The mean blackbody temperature is {\it kT}=0.093$\pm$0.007 keV with intrinsic dispersion $\sigma_{\it kT}$=0.034$\pm$0.005 keV. 
This value is in excellent agreement with recent estimates in the 
literature (e.g. Winter et al.~\cite{Winter09}; Bianchi et al.~\cite{Bianchi09}; Crummy et al.~\cite{Crummy06}, 
Piconcelli et al.~\cite{Piconcelli05}; Gierli\'nski \& Done~\cite{Gierlinski04}). We would like to note however that, although 
constraining the mean temperature of the soft excess using thermal models is a helpful measurement that can be compared with 
other data, it does not have any physical meaning.

As shown in Fig.~\ref{kT_prop} we find that the temperature of the thermal 
blackbody emission has no dependence on the redshift of the objects and is 
uncorrelated with the hard X-ray luminosity over more than two orders of magnitude in luminosity. 
This result agrees with the recent finding that the temperature of the soft excess is 
remarkably constant regardless of the luminosity and mass of the objects (e.g. Bianchi et al.~\cite{Bianchi09}; 
Crummy et al.~\cite{Crummy06}; Gierli\'nski \& Done~\cite{Gierlinski04}). 
Fig.~\ref{kT_prop} also shows the soft excess luminosity (rest-frame 0.5-2 keV, corrected for absorption) vs. the 
power-law X-ray luminosity (rest-frame 2-10 keV). If we use the ratio of the soft excess and power-law X-ray luminosities 
to measure the strength of the soft excess, we find that 
in most objects the strength of the soft excess is under 60\%.
Furthermore, there seems to be a strong correlation between the luminosities 
of the soft excess and power-law components, with ${\rm {\it log}(L^{se})=(1.2\pm0.1)\times {\it log}(L^{pow})-(9.2\pm 5.3)}$. This result 
can be explained as the combination of two different effects: first the 
lack of detections in the bottom-right of the plot in Fig.~\ref{kT_prop} (bottom-left) is due to the 
fact that it is more difficult to detect weak soft excess (strength lower than $\sim$10\%) on top of a bright continuum. Second, the lack of 
detections in the top-left of the plot suggests that there must be an upper limit in the strength of the soft excess that exists in AGN. The 
strongest soft excesses have been detected predominantly in narrow line Seyfert 1 galaxies (Boroson et al.~\cite{Boroson02}).

The measured soft excess blackbody temperatures are too high to be the high 
energy tail of thermal emission from the hot inner accretion disk ({\it kT}$\sim$20-40 eV for a black hole 
mass of ${\rm 10^7-10^8\,M\odot}$ accreting at the Eddington rate, Shakura \& Sunyaev~\cite{Shakura73}). 
Furthermore, the fact that the temperatures of the soft excess are 
$\sim$constant (0.1-0.2 keV) over such a broad range of luminosities and black hole masses 
is a major problem for models based on continuum emission, such as optically-thick Comptonization of accretion disk emission, because 
the disk temperature is expected to vary with both the mass of the black 
hole and the accretion rate (see Fig. 1c in Gierli\'nski \& Done~\cite{Gierlinski04}). However, see Kawaguchi et al.~(\cite{Kawaguchi01}) for a 
model in which optically-thin Comptonization is invoked to produce a soft excess 
with a shape that does not depend on the temperature of the disk.

Alternative models relate the soft excess to atomic rather than continuum processes which could easily explain the constancy 
of the temperature by the abrupt increase in opacity in partially ionised material between $\sim$0.7-3 keV especially due to 
${\rm O_{VII}/O_{VIII}}$ at 0.6-0.7 keV as well as Ly$\alpha$ lines from C, N and Fe L shell transitions (Ross \& Fabian~\cite{Ross05}). 
The result is a large increase in reflected/transmitted flux below 0.7 keV, which 
could produce the soft excess either from absorption in an optically thin material in the line of sight, for example in a wind from the 
disk (Gierli\'nski \& Done~\cite{Gierlinski04}), or from reflection in optically thick material out of the line of sight, for example 
in the accretion disk (Crummy et al.~\cite{Crummy06}). Both models predict sharp/strong atomic features from the partially ionised 
material. To explain the featureless soft excess they both require large velocity smearing which can only be produced in the 
vicinity of the black hole. In order to fit the strongest soft excesses detected these models require a fine tuning of the ionisation state.
Because the reflected emission cannot exceed the illuminating flux, this sets a tight limit on the strength of the soft excess in the reflection 
model, while in the absorption model the range of measured soft excess strength can be easily reproduced by changing the amount 
of absorption. Both models have been found to fit the X-ray spectral signatures of soft excess in bright AGN equally well. 

In an alternative scenario, it is possible that the soft excess emission does not come from the AGN but 
from the host galaxy (X-ray binaries, star formation, hot ionised gas). However 
this emission is not expected to be more luminous 
than ${\rm \sim10^{41}\,erg\,s^{-1}}$ (Ranalli et al.~\cite{Ranalli03}). The majority of the measured luminosities 
of the soft excess for our sources are at least an order of magnitude above this value, hence this component is 
expected to originate from the AGN in our objects.

\section{Conclusions}
We have analysed the 0.2-12 keV broad band spectral properties of one of the largest samples of X-ray selected type-1 AGN to date, drawn
from the XMM-{\it Newton} Wide Angle Survey. The sample contains 487 type-1 AGN detected up to redshift $\sim$4 and 
spanning more than 3 orders of magnitude in X-ray luminosity. The main results of our analysis are as follows.
   \begin{enumerate}
     \item The spectra of most type-1 AGN ($\sim$90\%) are best-fitted with a simple power-law. Absorption and soft excess emission 
       are detected with an F-test significance $>$99\% in 3\% and 7\% of the objects respectively.

      \item We constrain the mean spectral index of the broad band X-ray continuum of type-1 AGN to a value of 
	$\langle\Gamma\rangle=1.96\pm0.02$ with intrinsic dispersion 
	${\rm \sigma_{\langle \Gamma \rangle}=0.27_{-0.02}^{+0.01}}$, in agreement with previous estimates in the literature. 
	We find the continuum to strongly depend on the 0.5-2 keV flux, in the sense 
	that the type-1 AGN population becomes marginally harder at fainter fluxes. Our analysis 
	indicates that this spectral variation is likely due to undetected absorption in the faintest sources in our sample 
	rather than to changes in the intrinsic spectral shape.
	
      \item The mean continuum shape of our type-1 AGN is found to become harder at higher 2-10 keV luminosities and at higher redshifts. 
	At high redshifts undetected absorption is unlikely to explain the harder continuum of the objects as 
	we expect spectral signatures of moderate absorption to be redshifted outside the observed band-pass. The flattening of $\Gamma$ 
	could be the result of redshifting of the Compton reflection bump into the observed band and/or redshifting of the soft excess 
	outside the observed band. However, we find that there is a strong dependence of the detection efficiency of sources on the spectral 
	shape. Our analysis indicates that at high redshifts the efficiency of detection of hard objects increases while 
	for soft sources it decreases substantially. This effect must have an impact on the measured mean continuum shapes of sources at different redshifts
	and also at different luminosities due to the tight correlation between the X-ray luminosity and redshift in flux limited surveys.

      \item We find little or no evidence for X-ray absorption in the great majority of our type-1 AGN, as expected 
	from AGN unification models.
	However $\gtrsim$3\% of our objects show substantial absorption in X-rays.
	The measured amounts of X-ray absorption are typically lower than a ${\rm few\times10^{22}\,cm^{-2}}$, i.e. in the low part of 
	the column density distribution 
	of type-2 AGN. The recently presented clumpy torus models, where the torus is a smooth continuation of the broad line region, 
	can provide a physical explanation for the apparent mismatch between the optical classification and the X-ray properties of these objects. In these 
	models X-ray absorption, dust obscuration and broad line emission are produced in a single continuous distribution of clouds: the broad line region 
	is located inside the dust sublimation radius and hence the dust-free clouds obscure the X-rays but not the 
	optical while the torus is located outside the dust sublimation radius (dusty clouds). 
	
      \item A soft excess component is detected in 40 objects in our sample, which corresponds to 
	7.3\%$\pm$2\% of our type-1 AGN after taking into account spurious detections. 
	This value should be taken as a lower limit as for high redshift objects the signatures of soft excess will be redshifted 
	outside the observed band-pass. Indeed considering only objects at redshifts below 0.5, where most soft excess emission is detected, we found 
	that the fraction of objects with detected soft excess increases to $\sim$36\%.
	We constrain the properties of this spectral component with a thermal blackbody which provided the best quality fits to the data.
	We constrain the mean value of the temperature of the soft excess and the amplitude of the 
	intrinsic dispersion to {\it kT}$\sim$100 eV and $\sigma_{\it kT}$$\sim$34 eV. 
	We find that the temperature of the thermal blackbody emission has no dependence on the redshift of the objects and is 
	uncorrelated with the hard X-ray luminosity over more than two orders of magnitude in luminosity.
	The measured strength of the soft excess, defined as the ratio of the soft excess luminosity (rest-frame 0.5-2 keV, corrected for absorption) and the 
	power-law X-ray luminosity (rest-frame 2-10 keV), is under 60\% for the great majority of our objects.
	Direct thermal emission from the accretion disk is ruled out as a possible origin of the soft excess on the 
	basis of the very high temperatures detected for this component and the apparent lack of correlation of the temperature of the soft excess with  
	AGN parameters such as the X-ray luminosity. Compton scattered disk emission is also ruled out on the basis that in the Comptonization model 
	the shape of the soft excess should be related to the disk emission.
	The measured luminosities of the soft excess are at least an order of magnitude above those expected if the soft excess originates 
	in the host galaxy (X-ray binaries, star formation, hot ionised gas) hence this component is expected to originate from the AGN in our objects.
   \end{enumerate}

\begin{acknowledgements}
We acknowledge Chris Done, Bozena Czerny, Gordon Stewart, Pilar Esquej and Ken Pounds for useful comments.
SM, MW and JAT acknowledge support from the UK STFC research council. We acknowledge the anonymous referee for a careful reading of 
the manuscript and for comments that improved the paper. FJC acknowledges financial support for this work from the Spanish 
Ministerio de Educaci\'{o}n y Ciencia under project ESP2006-13608-C02-01. AC acknowledges financial support from the Spanish
Ministerio de Educaci\'{o}n y Ciencia fellowship and also from the
MIUR and The Italian Space Agency (ASI) grants PRIN$-$MUR 2006$-$02$-$5203 and n. I/088/06/0. MK acknowledges support from the NASA grant NNX08AX50G and NNX07AG02G.
\end{acknowledgements}

\end{document}